\documentclass{ws-ijmpa}

\newcommand{\beq}[1] {\begin{equation}\label{#1} }
\newcommand{\eeq} {\end{equation} }

\newcommand{\bea}[1]{\begin{eqnarray}\label{#1} }
\newcommand{\eea}{\end{eqnarray}}

\newcommand{\eps}{\epsilon}

\newcommand{\del}{\delta}

\newcommand{\lam}{\lambda}
\newcommand{\sig}{\sigma}

\newcommand{\der}{\partial}

\newcommand{\hmu}{{\hat\mu}}
\newcommand{\hnu}{{\hat\nu}}

\newcommand{\hh}{{\hat{h}}}

\newcommand{\tA}{{\widetilde{A}}}

\newcommand{\tih}{{\widetilde{h}}}
\newcommand{\tp}{{\widetilde\phi}}

\newcommand{\vn}{{\vec{n}}}

\begin{document}

\title{THE PHENOMENOLOGY OF UNIVERSAL EXTRA DIMENSIONS \\
AT HADRON COLLIDERS}
\author{COSMIN MACESANU}
\address{Department of Physics,
Syracuse University,\\
Syracuse, NY 13244-1130, USA\\
cmacesan@physics.syr.edu}

\maketitle

\begin{abstract}
Theories with extra dimensions of inverse TeV size (or larger) 
predict a multitude of signals which can be searched for at present and
future colliders. In this paper,
we review the different phenomenological signatures of a particular
class of models, 
universal extra dimensions, where all matter fields propagate in the bulk. 
Such models have interesting features, in particular Kaluza-Klein (KK)
number conservation, 
which makes their phenomenology 
similar to that of supersymmetric  theories.
Thus, KK excitations of matter are produced in pairs, and decay to a lightest
KK particle (LKP), which is stable and weakly interacting, and therefore will
appear as missing energy in the detector (similar to a neutralino LSP). 
Adding gravitational interactions which can break KK number conservation 
greatly expands the class of possible signatures. Thus, if gravity is the
primary cause for the  decay of KK 
excitations of matter, the experimental signals at hadron colliders
will be jets + missing energy,
which is typical of supergravity models. 
If the KK quarks and gluons decay first to the LKP, which then decays 
gravitationally,
the experimental signal will be photons and/or leptons (with some jets),
which resembles
the phenomenology of  gauge mediated supersymmetry breaking models.

\end{abstract}


\section{Introduction}

One of the deepest problems confronting our current understanding of
fundamental physics is the extreme weakness of the 
gravitational interaction compared with the other fundamental forces.
 Formulating this in terms of energy scales, the natural scale of 
the electroweak interactions is given by the mass of the $Z$ boson
$\sim 10^2$ GeV, while the natural scale of four-dimensonal gravity 
is given by the Planck mass $M_{Pl} \sim 10^{19}$ GeV.

This problem can be partially solved by assuming the existence of
extra spatial dimensions (besides the three we experience directly). 
The existence of extra dimensions is also consistent with the requirements
of string theory, which is considered by many a good candidate for the
theory of everything. `Classic' string  theory assumes that these extra
dimensions are compatified, with a radius of order $1/M_{Pl}$; this is the
reason why we do not observe them directly. However, this does not solve
the hierarchy problem.
Recent developements have allowed for new approaches to 
this problem.
 In the Randall-Sundrum type theories\cite{RS},
one assumes one extra dimension, but the metric of the space is anti-deSitter,
 rather than being flat.
As a consequence, gravity fields propagating in the fifth dimension suffer
exponential suppression. One then assumes that gravity lives on a different
brane than the Standard Model particles, and the weaknes of gravitational
interaction on our brane is due to gravitation having to propagate in this
fifth dimension.
 
 Conversely, in the Arkani-Hamed, Dimopoulos and Dvali (ADD) 
 approach\cite{ADD},
the extra dimensions are flat. The weakness of gravity is explained
by having the radius $r$ of the extra dimensions be rather large (of
order inverse eV, rather than inverse $M_{Pl}$). Then the strength of
gravitational interaction is diluted as it propagates on scales larger
than $r$. This is described by the relation
\beq{add_rel} M_{Pl}^2 \ \sim \ M_D^{N+2} r^N \ ,\eeq
where $N$ is the number of extra dimensions and $M_D$ is the Planck scale
in $4+N$ dimensions. One then sees that for values of the fundamental theory
parameter $M_D$  at  TeV scale, one will obtain the effective 4D Planck
mass for values of the radius $r$ ranging from $10^{-3} \hbox{eV}^{-1}$
(for $N=2$) to MeV$^{-1}$ (for $N=6$).
 
This picture leads to interesting consequences for low-energy 
phenomenology (here low meaning TeV scale). 
In the effective
4D theory, the 5D graviton field will appear as one 4D massless graviton 
plus an infinite number of 4D massive graviton fields (a Kaluza-Klein tower)
with masses equally spaced by an interval $\sim 1/r$. These individual
massive gravitons (also called Kaluza-Klein excitations) will each
have the same interaction with normal matter as the massless graviton,
whose strength is given by $1/M_{Pl}$. However, since their mass is so low,
this means that at energies reached  at current day colliders 
(few hundreds GeV)
one can produce a large  number of these KK excitations, with potentially
significant phenomenological consequences.

In the usual ADD scenario, the Standard Model (SM) matter fields (fermions and
bosons) are restricted to a 4D subspace called the SM brane. One could naturally
construct a more general theory by allowing the SM fields to also propagate
in the whole space (the bulk). However, this would imply that the SM particles
also aquire a KK tower of excitations with the same quantum numbers as the
original fields. Since one does not observe such excitations in colliders,
it means either that the SM fields do not propagate in the bulk, or that the
scale on which they propagate is much smaller (of order inverse TeV)
that the scale associated
with gravity, so that the masses of KK excitations of matter are
high enough to evade the experimental constraints.

Models in which only a subset of SM fields  (like, for example, the gauge
bosons) propagate in the bulk have been discussed for example in Refs. 
(\refcite{Accomando,Rizzo-lep,Nath-lhc,Dicus-lhc,Masip-mix,Rizzo-prec,Nath-prec}).
KK excitations of the gauge bosons can either be produced as final states
in colliders\cite{Accomando,Nath-lhc,Dicus-lhc,Rizzo-lep}, 
or they can affect precision electroweak observables 
(like the Z boson couplings, or the $\rho$ parameter) either
through mixing with the SM gauge bosons\cite{Rizzo-lep,Masip-mix} or through radiative
corrections\cite{Rizzo-prec,Nath-prec}.
For such
models, it has been found that electroweak precision data impose quite
stringent constraints, requiring for example that the compactification
scale for the space on which the SM fields propagate be larger than several
TeV.

We shall consider in this paper a set of models in which all matter fields
propagate in one extra-dimension with a radius $R \sim$ TeV$^{-1}$. These
models are called Universal Extra Dimensions (UED) models, and their
phenomenology has been studied initially in Refs. (\refcite{acd,Rizzo-ued}).
Compared with theories where
only the gauge bosons propagate in the bulk, UED models have some specific
features. The most important one is conservation of momentum in the
extra dimensions; this will result in a selection rule called 
Kaluza Klein number conservation. As a consequence, at tree level 
one can only couple an even number of KK excitations to a SM field.
This means that one has to pair produce these KK excitations at colliders.
Also, the SM fields do not mix with the first level KK excitations, and
radiative corrections to any electroweak precision observable are suppressed
by the requirement of having two massive particles in the loop. These 
properties imply that experimental constraints on UED models are quite weak; 
both present day collider data\cite{Rizzo-ued,cmn} and 
electroweak precision measurements\cite{acd,Rizzo-ued,Agashe-bsg,dobrescu-mu}
can be satisfied with
a compactification scale of the order of 300 GeV\footnote{Newer analysis
of electroweak precision data\cite{hooper} 
sets a limit on the masses of first level
KK excitations as high as 700 GeV}.

This implies that such theories may have interesting predictions for collider
phenomenology. The mass range accesible for UED models 
at present and future colliders is similar to that of supersymmetric (SUSY)
models.
One can actually draw many parallels between SUSY and UED phenomenology 
(as has been initially noted in Ref. (\refcite{CMSexp})). 
Both theories imply the existence of a sector of heavy particles
with the same  couplings and quantum numbers as the SM ones; only that 
in the case of SUSY the partners of the SM particles have different spins, 
while in UED they have the same spin. (In the UED case, one will have actually
a whole tower of such states, but since the masses are rather high,
for purposes of phenomenology only the first level will probably be 
important).
Moreover, the KK number conservation rule in UED models is similar to R-parity
in SUSY; both these selection rules imply that the lightest heavy partner
is stable, which makes is a good candidate for dark matter.
 An obvious difference between the two theories 
is that in SUSY models the masses of the superpartners
are generally non-degenerate; while for UED models, the masses of the first
level KK excitations are more or less the same, at least at tree level.
However, radiative corrections can lift this degeneracy up to a certain 
point\cite{CMS}.

Assuming that  KK excitations of SM matter can be produced at colliders, 
in order to observe them one must know first if they decay, and if yes what
are the relevant modes.
At tree level, the masses of the lowest level KK excitations are almost
degenerated (the splitting between them is provided by the SM mass terms,
and therefore is extremely small for all particles with the exception of
the top quark or the heavy gauge bosons). This would imply that most first
level KK particles are stable, and therefore very hard to see at colliders
(the parameters of the model will be however strongly constrained by 
cosmological restrictions on heavy stable charged particles). Going
beyond tree level, one however finds that loop corrections are able to 
lift this degeneracy. Then only the lowest mass KK excitation (which
most probably will be the photon partner $\gamma^*$)
 will be stable, providing an
excellent candidate for dark matter\cite{Servant-dm,Feng-dm}.
 The other KK excitations
produced in colliders will cascade decay to $\gamma^*$, radiating SM 
particles in the process. However, these particles will be rather soft,
since the total energy available to them 
is equal to the mass splitting between the particle initiating the cascade
and the $\gamma^*$, which is typically not very big. As a consequence, it will
be difficult to see such decays at hadron colliders; lepton colliders would
provide a better place to test this type of models\cite{CMSexp}.

One could consider an alternate type of models, where KK number conservation
is broken by a new interaction, thus leading to the direct decay of 
the first level KK excitations. Such an interaction can be provided in a 
natural way by gravity. As metioned above, in ADD type models the natural
scale for the propagation of gravity in extra dimensions is of order
inverse eV, which is many orders of magnitude larger than the scale on
which the SM matter fields propagate in extra dimensions in UED type models. 
One could then consider a fat brane scenario\cite{DeRujula:2000he}:
there are N extra dimensions
of inverse eV size, into which gravity propagates. However, matter propagation
is restricted only to a small length  (of order inverse TeV) - associated
with the thickness of the brane - along these
extra dimensions. One has the added benefit that the energy scale associated
with matter propagation would be close to the fundamental gravity scale
$M_D$; one could imagine then that there is some physics at $M_D$ scale
which confines the matter fields close to the 4D brane. 

From a phenomenological point of view, this scenario also has 
interesting consequences. KK excitations produced at colliders can decay
directly to their SM partners by radiating a graviton. The experimental 
signal\cite{Rizzo-ued,cmn} would be two jets
 with large $p_T$ (since they come from the decay
of a heavy particle) plus missing energy (taken away by the gravitons).
An even more interesting scenario can appear in the case when both decays due
to mass splitting and due to gravity interaction take place. Then, one
could have for example the excitation of a quark decay to $\gamma^*$ through
usual electroweak interaction, and consequently the  $\gamma^*$
decay to a photon and a graviton through gravity mediated interaction.
The experimental signal for this case will be quite striking at a hadron
collider, consisting of two large $p_T$ photons plus missing energy\cite{cmn2}.

Adding gravitational interactions has additional implications for
the production of KK matter states. Since the coupling of matter to gravitons
does not obey  KK number conservation rules, it is possible to
produce  single KK excitations of quarks or gluons at  colliders. This would 
mean that one can probe higher values for the compactification scale 
associated with matter fields (since there is no need to have two 
heavy particles in the final state). There are
two types of new processes mediated by gravity:
one in which gravitons act as virtual particles appearing as the $s,t$ and $u$
channels propagators in amplitudes describing the  
production of a SM particle and a KK excitation\cite{marius1}, and other in
which gravitons appear in the final state together with 
a KK particle\cite{Marius2}. For
the first case the collider signal will be two jets plus missing energy (or
a jet plus photon/lepton with missing energy, if the KK excitation decays
first to the LKP), while for the second case the signal
will be a single jet/photon/lepton plus missing energy (again
depending on the decay pattern of the KK particle produced).

The outline of this paper is as follows. In the next section we will start by
reviewing the basic formulation of a theory with 
extra dimensions as an effective
4D theory in terms of KK excitations. We will explain what is 
the particle content of the theory, as well as how to derive the interaction
lagrangian. The production rates for a pair of 
KK excitations at the Tevatron and LHC will be presented.
We will then review the
issue of radiative corrections to the tree level lagrangian, and how these
affect the masses of the first level KK excitations. We continue by discussing
the decay modes of KK excitations of gluons and quarks,
 and the corresponding phenomenological signal at hadron colliders.

In section 3 we will introduce gravity in our model. First, we review the
decomposition of $4+N$ dimensions gravity in KK modes, and the derivation
of gravity matter interactions in the fat-brane scenario. The gravitational
decay widths of the KK matter excitations are computed. We follow by discussing
the phenomenology of matter KK pair production with decays mediated by gravity
(having either the initial KK excitations of quarks and gluons 
decay through graviton radiation, or
the LKP decay through graviton radiation). Then, we discuss the 
gravity-mediated production of a single matter KK excitation. 
We end with conclusions.

\section{Matter in extra dimensions}

This section will be devoted to a discussion of the  features of
models where all matter fields propagate in extra dimensions. For simplicity,
we will consider in detail the case of one extra dimension only; models
with more than one ED can have some interesting properties
from a theoretical perspective\cite{Dob01a,Dob01b},
but the phenomenology will be quite similar\cite{Dob05}.
We will start with the derivation 
of the 4D effective Lagrangian, evaluate the production rates at the Tevatron
and LHC, and discuss the one loop radiative corrections and their effects
on the decay modes on the first level KK excitations.

\subsection{The effective 4D description}

Consider a theory defined on a five-dimensional extension of the Minkowski
 manifold. Then, a general field
$\phi$ will depend on the usual Minkowski coordinates $x_0, \ldots , x_3$
as well as on the extra coordinate $y$. Assuming that the extra dimension
is compactified as a circle of radius $R$,
one can write $\phi(x,y) = \phi(x,y+ 2\pi R)$. Then the fields defined
on this manifold can be expanded in a Fourier series along the $y$
coordinate:
\beq{exp1}
\phi(x,y) \ =\ \frac{1}{\sqrt{2\pi R}}\left[ \phi_0 (x) +
\sqrt{2} \sum_{n=1}^{\infty} \left( \phi^{(1)}_n (x) \cos(\frac{n y}{R})
+ \phi^{(2)}_n (x) \sin(\frac{n y}{R})
\right) \right] \ .
\eeq
Here $\phi_0 (x),\ \phi^{(1)}_n (x)$ and $ \phi^{(2)}_n(x)$ are 4D fields; 
in the effective theory framework,
$\phi_0$ would be indentified with the Standard Model field, while 
$\phi^{(1)}_n$ and $ \phi^{(2)}_n$ would be its KK excitations. The constants 
in front are chosen so one has the normalization:
$$\langle \phi(x,y) , \phi(x,y) \rangle \ = \ 
\langle \phi_0(x), \phi_0(x) \rangle
\ + \ \sum_{n=1}^{\infty} \left( 
\langle \phi^{(1)}_n(x), \phi^{(1)}_n(x) \rangle 
\ + \ \langle \phi^{(2)}_n(x),\phi^{(2)}_n(x) \rangle \right) \  .
$$

Assuming that $\phi$ stands for a 5D scalar field with the Lagrangian
density
$$ {\cal L}_5(x,y) \ = \ \der_M \phi \ \der^M \phi 
\ = \ \der_{\mu} \phi \ \der^{\mu} \phi \ + \ (\der_y \phi)^2 $$
(with $M$ taking values 0, \ldots , 3 and 5),
upon expansion in KK modes (\ref{exp1}) and integrating over the fifth
coordinate $y$, one would get an effective Lagrangian
\bea{eff_scal}
{\cal L}_{eff} & = & \der_{\mu} \phi_0 \ \der^{\mu} \phi_0 \ + \
\sum_{n=1}^{\infty} \left[ 
\der_{\mu} \phi^{(1)}_n \ \der^{\mu} \phi^{(1)}_n \ + \ 
\left( \frac{n}{R} \right)^2 (\phi^{(1)}_n)^2
\right] \nonumber \\
& & + \sum_{n=1}^{\infty} \left[ 
\der_{\mu} \phi^{(2)}_n \ \der^{\mu} \phi^{(2)}_n \ + \ 
\left( \frac{n}{R} \right)^2 (\phi^{(2)}_n)^2
\right] \ .
\eea
We see that zero mode stays massless, while each KK excitation $\phi_n$
gets mass equal to $n/R$.

One notes that at each KK level, there are twice the degrees of freedom
that appear at the zero level. It is possible to eliminate some of these
degrees of freedom by imposing extra symmetries acting on the fifth
coordinate. For example, one could require that the fields appearing in
the theory have some definite property under the reflection $y \rightarrow -y$.
If we require that the fields are even $\phi(x,y) = \phi(x,-y)$, then the
coefficients $\phi^{(2)}_n$ of the sine modes should be set to zero; conversely,
requiring that the fields are odd $\phi(x,y) = -\phi(x,-y)$
eliminates the coefficients $\phi^{(1)}_n$ of the cosine functions, as well
as the zero mode.
 
Imposing these types of symmetries goes by the name of orbifolding. That
is, one replaces the circle $S_1$ by the quotient space (orbifold) $S_1/Z_2$
constructed by identifying $y$ with $-y$ (folding the circle upon itself).
The rationale for introducing such constructions is to eliminate from
our theory
extra degrees of freedom appearing at the zero level. For example, consider
a 5D massless gauge vector field. 
In five dimensions, it will have five components $(A_{\mu},
A_5)$; we take $\mu$ to stand for the usual Minkowski indices $0,\ldots , 3$. 
One can expand each component into a Fourier series as in (\ref{exp1}). It 
can then be shown that at zero level the five components $(A_0)_{ \mu},
(A_{0})_5$ stay massless, while at each KK level there exists a particular
gauge choice such as the four fields  $(A_n)_{ \mu}$ became
the components of a 4D  massive gauge boson (with mass $n/R$),
while the $(A_n)_5$ component disappears from the theory\footnote{
The disappearance of the $(A_n)_5$ component can be understood remembering
that in 4D a massive gauge boson has one extra degree of freedom compared to a 
massless one. Thus the $(A_n^{(1,2)})_{ \mu}$ bosons get mass by absorbing the 
$(A_n^{(1,2)})_5$ fields, which therefore play a role similar to that of
Goldstone bosons in the Higgs mechanism. For a more detailed discussion 
see, for example Refs. (\refcite{Dienes,Santamaria})}.
One is therefore left with an 
extra massless boson in the effective 4D theory - the $(A_0)_5$ - which
will be a scalar under the Lorentz transformations associated with 4D Minkowski
space. This boson will have the same interactions as the gauge fields, 
therefore giving rise to obvious phenomenological problems. However,
one can make use by the orbifold construction to eliminate some degrees
of freedom. Thus, requiring that the $A_{\mu}$ components are even
under the transformation $y \rightarrow -y$ will eliminate the odd modes,
leaving the even ones including the zero order modes, which will
play the role of the SM gauge fields. Conversely, requiring that 
$A_5$ field is odd\footnote{Note that invariance of the term $\der^M A_M$
under reflection with respect to the $y$ coordinate implies opposite
transformation properties for the $A_{\mu}$ and $A_5$ fields.}
under the orbifold transformation will eliminate the even modes, which
include the zero order mode.

We will also take the scalar fields (the Higgs multiplet only, for the SM case)
to be even under the orbifold symmetry.
With these properties, the 5D scalar and gauge fields
 have the following decomposition in KK modes 
\beq{sm_exp} (\Phi, B_{\mu}^a)  \ = \
\frac{1}{\sqrt{\pi R}}\left[ (\Phi_0, B_{\mu, 0}^a)  +
\sqrt{2} \sum_{n=1}^{\infty} (\Phi_n, B_{\mu ,n}^{a})  \cos(\frac{n y}{R}) \right] \ . \eeq
Here $B$ stands for the $U(1)_Y$, $SU(2)_L$ or $SU(3)_C$ gauge fields ($a$
being the coresponding Lie algebra index). Since the fields are defined on
an orbifold, we take the integration range for the $y$ variable from
zero to $\pi R$; hence the normalization factor in front of the expansion is
$\sqrt{\pi R}$.

Let us now consider the fermion sector. As it is well known, there are
some subtleties involved with defining fermions in more than four dimensions.
Fortunatelly, the case of five dimensions is quite simple. A 5D fermion
field can still be expanded in terms of the 4 component Dirac spinors
we are familiar with from four dimensions\footnote{For more than five 
dimensions, the matrix representation of the $\Gamma_M$ generators
of the Clifford algebra has to be necessarily in more than four dimensions;
for example in 6D the $\Gamma_M$ will be $8\times 8$ matrices, hence
a 6D spinor will have eight components.}. Decomposing in Fourier modes 
on a circle, one can write
\beq{exp1_fer}
\psi(x,y) \ =\ \frac{1}{\sqrt{2\pi R}}\left[ \psi_0 (x) +
\sqrt{2} \sum_{n=1}^{\infty} \left( \psi^{(1)}_n (x) \cos(\frac{n y}{R})
+ \psi^{(2)}_n (x) \sin(\frac{n y}{R})
\right) \right] \ .
\eeq
If $ \psi(x,y) $ satisfies the 5D Dirac equation 
$\der_M \Gamma^M  \psi(x,y) =0 $, with $\Gamma^{\mu} = \gamma^{\mu}$, 
$ \Gamma^5 = i\gamma_5 = \gamma^0 \gamma^1 \gamma^2 \gamma^3$, and
$\gamma^{\mu}$ the usual 4D Dirac matrices, one has
$$
\der_{\mu} \gamma^\mu  \ \psi_0(x ) \ = \ 0 \  , \ \ \ \
\der_{\mu} \gamma^\mu  \ \psi^{(1,2)}_n(x) \ = \
 \pm \frac{n}{R} \psi^{(1,2)}(x) \ ,
 $$
hence $\psi^{(1,2)}_n$ are 4D spinors with mass $n/R$. 

The only subtlety 
involved with the 5D fermions has to do with the fact that there is 
no chirality in five dimensions. This is due to the fact that one cannot 
construct a matrix with the properties of $\gamma_5$ in 4D; that is, 
anticommutes with all $\Gamma^M$, and its square is identity. What this
means  from a practical point of view 
is that bilinears like $\psi \gamma^\mu \gamma_5 \psi$ are not invariant under
5D Lorenz transformations, so they cannot appear in the 5D Lagrangiang.
As a consequence, one cannot have the left and right handed components
of the zero excitation $\psi_0$ couple differently to the gauge
fields as in the Standard Model. So one cannot get a chiral SM fermion by using
a single 5D fermion field.

It is therefore necessary that for each Dirac fermion field $\psi^{SM}$
appearing in the Standard Model we introduce two 5D fermion fields:
one field $\psi$ which has the quantum numbers of the 
left handed $\psi^{SM}_L$ spinor, and
one field $\psi'$ with the quantum numbers of the right handed $\psi^{SM}_R$
spinor. One can then set $\psi^{SM} = P_L \psi_0 + P_R \psi'_0$, with
$P_R, P_L$ the chirality projectors $(1 \pm \gamma_5)/2$. However,
we are then again left with extra massless degrees of freedom at the zero level,
specifically the right handed components of $\psi_0$ and left handed
components of $\psi'_0$.

As for the case of gauge boson fields, formulating the theory on an orbifold
will help us eliminate these degrees of freedom. We will thus require that
$\psi$ is odd under the $y \rightarrow -y$ orbifold symmetry (under
which the spinor fields transform as 
$\psi(x,y) \rightarrow \gamma_5 \psi(x,-y)$), while $\psi'$ is even.
One can then write the decomposition in terms of the remaining modes
\bea{sm_fer}
\psi  &  = &
\frac{1}{\sqrt{\pi R}} \left\{ \psi_{0L} + \sqrt{2} \sum_{n=1}^{\infty}
\left[ \psi_{nL}^{(1)}  \cos \left(\frac{n y}{R} \right)
 + \psi_{nR}^{(2)}  \sin \left(\frac{n y}{R} \right) \right] \right\} \nonumber \\
\psi'  &  = &
\frac{1}{\sqrt{\pi R}} \left\{ \psi'_{0R} + \sqrt{2} \sum_{n=1}^{\infty}
\left[ \psi_{nR}^{\prime(1)}  \cos \left(\frac{n y}{R} \right)
 + \psi_{nL}^{\prime(2)}  \sin \left(\frac{n y}{R} \right) \right] \right\} \ ,
\eea
where $\psi_{L,R}^{(1,2)} = P_{L,R}\ \psi^{(1,2)}$. At the zero level one
then gets the SM field 
$\psi^{SM} = \psi_{0L} + \psi'_{0R}$. At each $n$ level one gets two KK
excitations $\psi_n = \psi_{nL}^{(1)} + \psi_{nR}^{(2)}$ and
$\psi'_n = \psi_{nR}^{\prime(1)} - \psi_{nL}^{\prime(2)}$ with KK mass $n/R$
(the minus sign in the definition of $\psi'_n$ is chosen so that the mass 
term has
proper sign). Moreover, the $\psi_n$ excitations will have the same
quantum numbers as the left handed components of the SM fermion $\psi^{SM}$
(and we will call them left-type excitations), while the 
the quantum numbers of $\psi'_n$ excitations will be the same as
those of the right handed components of $\psi^{SM}$ (and we will call 
them right-type excitations).

 Once the field content of the theory had been determined, one can
write directly the lagrangian of the theory:
\bea{lag_gen}
{\cal L} & = & - \frac{1}{4} \sum_B F_{\mu \nu} F^{\mu \nu} \ + \
\sum_{\psi}i  \bar{\psi} \Gamma^M D_M \psi
\ + \ (D^M \Phi)^\dag (D_M \Phi) 
\ + \ 
\nonumber \\
& & + \ \sum_{\psi,\psi'} \lam_5 \bar{\psi} \Phi \psi' \ - \ V(\Phi) \ .
\eea
The first line in the above expression
contains the kinetic terms for the gauge, fermion and 
Higgs fields (with sums over all the gauge and fermion fields), 
while the second line contains the Higgs-fermion coupling
terms and the Higgs potential. (For the Standard Model case, $\Phi$ stands
for a single Higgs $SU(2)_L$ doublet.) 
The derivatives $D_M$ are the proper
covariant derivatives $D_M = \der_M - i g_5 \sum_a T^a B^a_M$,
$T^a$ being the Lie algebra generators.
One obtains the effective 4D lagrangian by expanding the fields in KK
modes and integrating over the $y$ coordinate.

From the effective lagrangian, one can work out the Feynman rules
governing the interactions of the KK excitations among
themselves as well as with the SM particles (some detailed discussion
can be found in, for example,
Ref. (\refcite{cmn})). We will comment on some interesting aspects of the model.
For example, note that the coupling constants appearing in the 5D theory
have dimensions of (mass)$^{-1/2}$ (in a theory with $N$ extra dimensions,
they would have dimension of (mass)$^{-N/2}$). They are related to 
the four dimension coupling constants by $g_5 = g/\sqrt{\pi R}$
(similar to the ADD relation (\ref{add_rel})). This also means that the theory
is nonrenormalizable. From the 4D effective theory viewpoint, 
nonrenormalizability is a consequence of the fact that there are
an infinite numbers of degrees of freedom in the theory. However, for
phenomenological purposes, one can truncate the theory to a small number
of modes, and such a truncated theory is renormalizable. 

Let us now consider interactions involving one or more KK excitations.
The terms in the Lagrangian describing such interactions will 
generally be trilinear in
a mix of the gauge, fermions and Higgs fields. For example, the gauge 
interactions of fermions will be derived from
\bea{ffa_int} 
{\cal L}_{\psi \psi B} & = & \int_0^{\pi R} d y \ 
g_5 \ (\bar{\psi } \gamma^M  T^a B^a_M  \psi) (x,y)  \
 \subset \ g_5 \sum_{m,n,k \ge 0} \bar{\psi}_n \gamma^\mu T^a \psi_m
B^{a}_{\mu,k} \nonumber \\
& & \times \int_0^{\pi R} \frac{dy (\sqrt{2})^{\theta(m)+\theta(n)+\theta(k)} }
{(\pi R)^{3/2}}  
\cos \left(\frac{m y}{R} \right)
\cos \left(\frac{n y}{R} \right)
\cos \left(\frac{k y}{R} \right)   \ , 
\eea
with $\theta (i) = 0$ for $i=0$ and 1 otherwise.
The condition that the integral over $y$ is nonzero imposes the KK conservation
rule
$$ |m \pm n \pm k | \ = \ 0 \ ,$$ 
which requires that, for example, if a SM particle participate in the 
interaction, the other two have to be KK excitations with the same KK number.
From expression (\ref{ffa_int}) one can also see that the coupling constant
appearing in the vertex will be $g = g_5/\sqrt{\pi R}$ if one of the
particles is a Standard Model one ($m,n$ or $k$ is zero), or
$g/\sqrt{2}$ if all three are massive KK excitations. For more details see
Ref. (\refcite{cmn}). 

Finally, we make some comments on the masses of the KK fermion excitations.
These particles get mass from two sources. One is the $\sim 1/R$ contribution
coming from the extra dimensions; the other is the Higgs mechanism (we need
that the SM Higgs field aquire a vacuum expectation value in 
order to give mass to the Standard Model fermions). However, as can be seen
from Eq. (\ref{lag_gen}), the Higgs couples the left-handed fields $\psi$
to the right-handed ones $\psi'$; the mass matrix for the fermion excitations
at a KK level $n$ will then have an off-diagonal structure
$$ {\cal L}_{mass}^n \ = \ 
\left( \begin{array}{cc} \bar{\psi}_n & \bar{\psi}_n' \end{array}  \right)
\left( \begin{array}{cc} \frac{n}{R} & \lambda \langle h_0 \rangle \\
        \lambda \langle h_0 \rangle & \frac{n}{R} \end{array}  \right)
\left( \begin{array}{c} \psi_n \\ \psi_n'  \end{array}  \right)
$$
where $ \lambda \langle h_0 \rangle = m_{SM}$ is the mass of the SM fermion.
The mass eigenstates for the fermion excitations will then contain an admixture
of both right and left-handed fields; however, since the amount of mixing
is $\sim m_{SM}/(n/R)$, we will generally neglect this (it could be important
for the top quark fields, but top does not play a big role in our analysis). In
keeping with this approximation,
we will also take the tree level mass of the KK excitations
to be given by $n/R$.

\subsection{Pair production of KK excitations}

In models with universal extra dimensions described in the previous section,
the main way of obtaining KK matter excitations will be pair production
at hadron colliders. Since the couplings of KK matter are the same as those
of their SM partners, one will produce mostly the first
level excitations of gluons and quarks. The amplitude
squared for the production process will be of order $\alpha_s^2$ (as in the
case of SM matter) but the
cross-section will be suppressed by the kinematics of the process. Since
one has to produce two massive particles in the final state, one needs
a large center-of-mass (CM) energy. 
Hence, production of such excitations at $e^+ e^-$ colliders would require
that the scale of extra dimensions be rather large.

\begin{figure}[b!] 
\centerline{
   \includegraphics[height=2.5in]{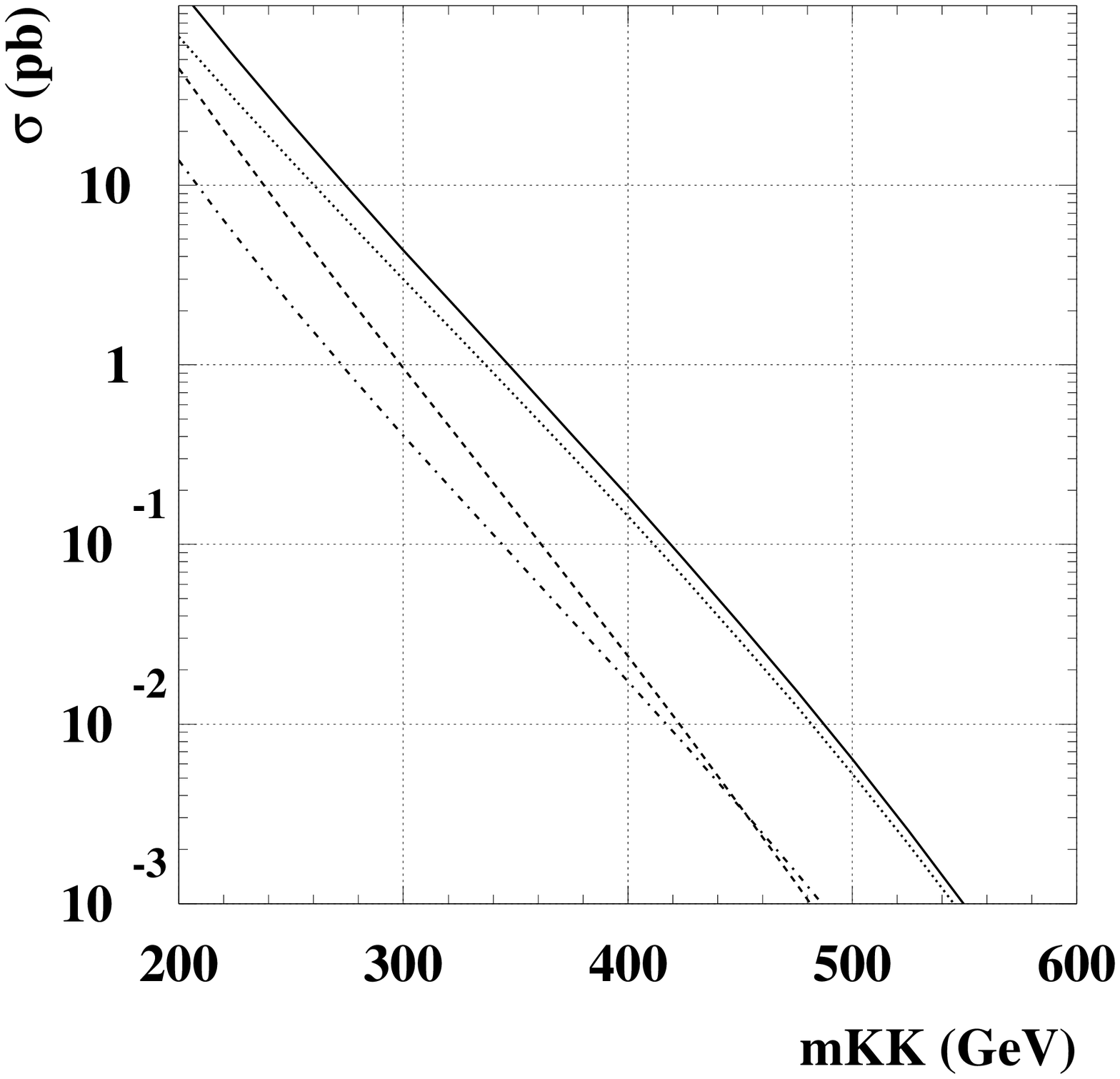}
   \includegraphics[height=2.5in]{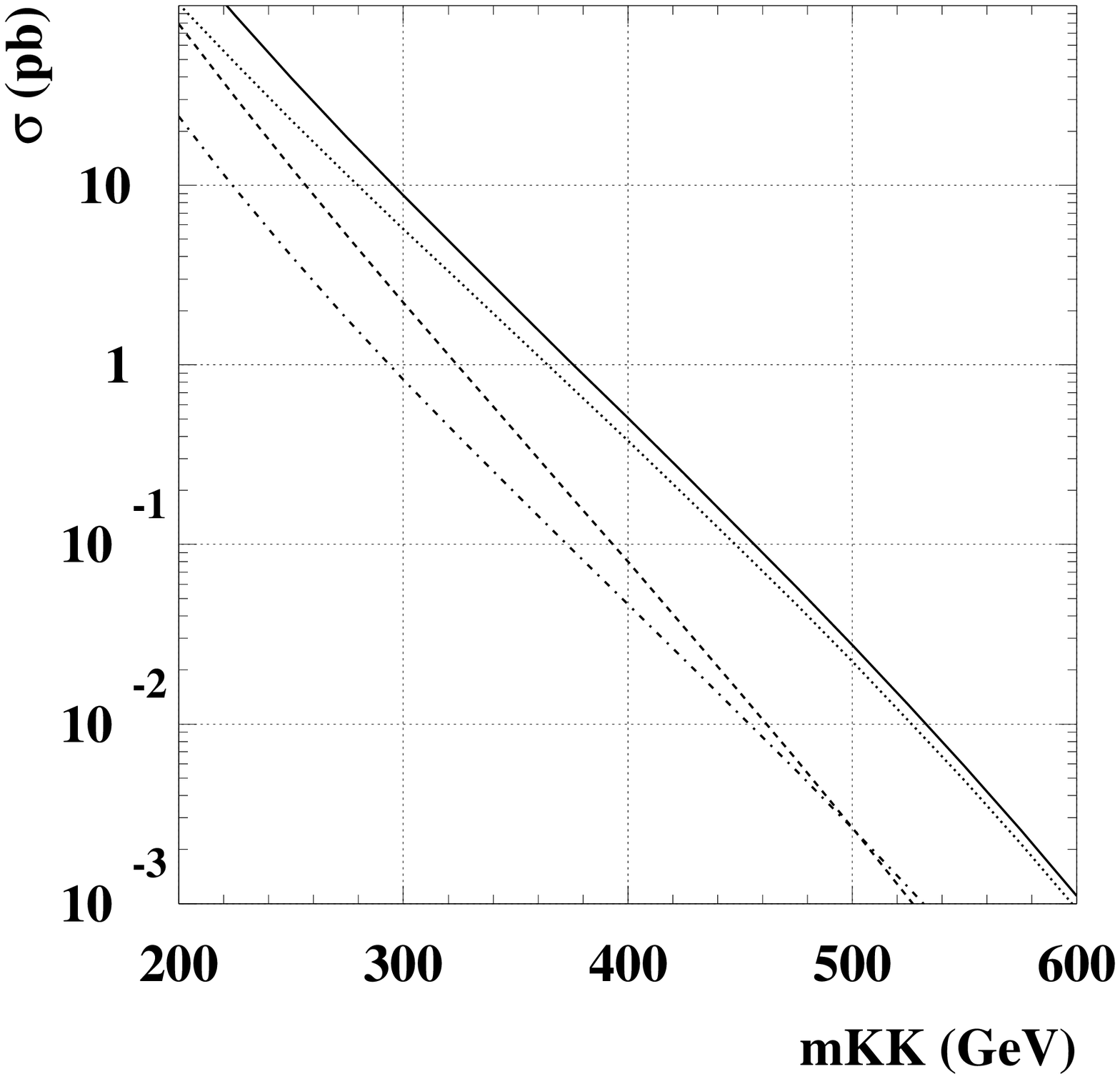}
   }
\caption{Tevatron Run I (left) and Run II (right) production rates
for KK pairs. 
(The solid line is the total cross section, while the dotted, dashed
and dot-dashed lines correspond to different types of final states, as
described in the text). }
\label{tev_KK_prod}
\end{figure}

Following the discussion in Ref. (\refcite{cmn}), we review in this section the production rates for KK pairs at the Tevatron
and Large Hadron Collider (LHC). The same processes contribute to the final
state with two KK particles in both cases. These processes can be clasified
after the particle content of the final state: processes with two quark
KK excitations ($a$), one quark excitation and one gluon excitation ($b$),
or two KK gluons ($c$). In Fig. \ref{tev_KK_prod} we show the production
rates for type $a$ processes (dotted line), type $b$ (dashed line), type $c$
(dot-dashed line), as well as for their sum  (the solid line), which is the
total cross section for the production
of two KK excitations. The left plot corresponds to Run I case (center
of mass energy 1.8 TeV), while the right plot corresponds to Run II case
(CM energy 2 TeV). One can see that at Run I, with an integrated luminosity
of 100 pb$^{-1}$, one can produce at least some KK excitations in this 
scenario if the mass scale associated with the radius of 
the one extra dimension where matter propagate is
$m_{KK} = 1/R \lesssim$ 500 GeV. At the Tevatron Run II, with an integrated
luminosity of order 1 fb$^{-1}$, one can produce KK excitations if 
$1/R \lesssim$ 600 GeV.

The question arises then what will happen with the first level KK excitations
once they are produced. As mentioned in the previous section, KK number
conservation dictates that, if these particles are degenerate in mass, they
will be stable. However, as we will be showing in the following, loop
corrections  break the degeneracy, allowing the heavier first level KK
excitation to decay to the lightest one.  
Furthermore, there are mechanisms which can break the 
KK number conservation rules;
for example, adding boundary terms (such terms can also be generated naturally
by loop effects) or considering new interactions (for example, 
gravitational interaction). The phenomenology arising from such scenarios will 
be discussed in the next sections.

\begin{figure}[t!] 
\centerline{
   \includegraphics[height=2.5in]{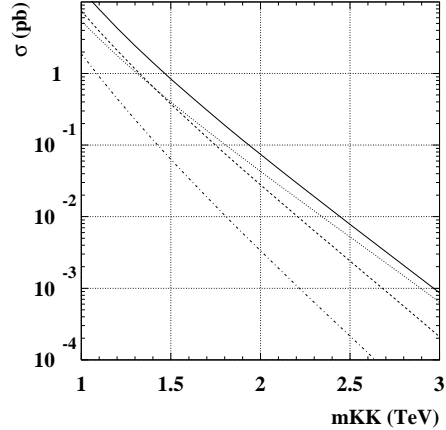}
   }
\caption{LHC production rates for KK pairs. 
(The solid line is the total cross section, while the dotted, dashed
and dot-dashed lines correspond to different types of final states, as 
described in the text). }
\label{lhc_KK_prod}
\end{figure} 

Here we will shortly discuss the case when the KK excitations of quarks and
gluons produced at a hadron collider live long enough that they do not decay 
in the detector. Such particles will hadronize while moving through the 
detector, producing high-ionization tracks. Their signal will be identical to
that produced by heavy stable quarks. Searches for such particles\cite{connoly}
 at the Tevatron Run I have already set an 
upper limit of around 1 pb on the
production cross section, although for particles somewhat lighter than those 
considered here. Using this limit, one can set an estimate lower bound on
the mass of long lived KK excitations of around 350 GeV, comparable with
the limit one obtains from electroweak precision data\cite{acd}.

We show in Fig. \ref{lhc_KK_prod} the production rates for first level
KK excitations of quarks and gluons at the Large Hadron Collider (LHC). 
The solid line is the 
total production cross section, while the dotted, dashed and dot-dashed line
correspond to production of two KK quarks, one KK quark and one KK gluon, and
two KK gluons respectively. One can see that the potential reach of the LHC
collider with an integrated luminosity of 100 fb$^{-1}$ extends to above 3 TeV
for the mass of one KK excitation. However, for a precise analysis one needs 
to know the decay patterns of the KK particles. This will be discussed in the
following.

\subsection{One-loop corrections}

The equality of KK particle masses at  particular KK level is a tree level
relation. Such relations, unless protected by symmetries, are generally
breaked by radiative corrections. So, on general grounds, one would
expect that the first level KK quarks, for example, would aquire a 
one-loop mass correction of order
$$ \del_{m_q} \ \sim \ \frac{1}{R} \ \frac{g_s^2}{16\pi^2} \times  {\cal O}(1)
\ . $$
Here, $g_s$ is the strong
interaction coupling constant, since the dominant correction for
quarks will be due to
gluon loops, while $16 \pi^2$ is a loop factor. Conversely, for the 
KK excitation of a lepton, 
one would expect that the mass correction be proportional to $g^2/(16\pi^2)$,
where $g$ is the electroweak coupling constant. We see then that
one might expect that the masses of 
KK excitations at each level will aquire splittings of order
percent of the tree level mass,
hence potentially allowing a heavier (strongly interacting) 
KK excitation to decay to 
a lighter one (with only electroweak interactions) 
with the same KK number.

The quantitative analysis of the one loop effective lagrangian can be found in
Ref. (\refcite{CMS}). We summarize the results here. There are two types of
contributions to the masses of KK particles at one loop. First type can be
thought to arise from the propagation of gauge fields in the fifth dimension 
(or as corrections due to KK excitations of fields), and are called bulk
terms.  The second type are due to the orbifolding conditions. These
conditions break translational invariance at the orbifold boundaries
($y = 0$ and $y = \pi R$); as a consequence, radiative corrections
generate boundary localized terms which break KK number 
conservation\cite{hailu}. For
example, contributions to the renormalized lagrangian arising from corrections
to the fermion propagators will look like:
$$ \del {\cal L}_{\phi \bar{\phi}} \ \sim \ \frac{\del(y) + \del(y - \pi R)}{2}
\frac{g^2}{64 \pi^2} \ln \left( \frac{\Lambda^2}{\mu^2} \right)
\left[ \bar{\psi}_L i \not{\der} \psi_L \ + \
	(\der_5 \bar{\psi}_R) \psi_L  \ + \ 
\bar{\psi}_L  (\der_5 \psi_R)\right] \ .
$$
The right-handed fields vanish on the boundary, so the kinetic energy terms
for these do not get corrections. 
Note, moreover, that these terms are log divergent, and one has to specify some 
renormalization conditions. In the above expression, this condition is
that at some large scale $\Lambda$ the boundary terms are zero. As a 
consequence, the boundary corrections to the KK excitation masses
are enhanced by a factor $\ln (\Lambda^2/\mu^2)$, where $\mu$ is the scale 
relevant for the process under consideration (for example, for production
of KK excitations at colliders one can take $\mu = m_{KK}$).

  Taking into account radiative corrections introduces therefore a new parameter
in the model: the cut-off scale $\Lambda$. This can be thought of as the
energy scale up to which the effective description of the theory in terms
of 4D KK excitations works. Lacking information on the parameters of the
underlying fundamental theory, we no not know what this scale is. However, one 
can make educated guesses. For example, one can use unitarity bounds on 
heavy gluon scattering to put limits on the maximum number of KK excitations
which can appear in the effective 4D theory\cite{Dicus-uni}. Such bounds
typically show that one cannot have more then ${\cal O}(10)$ KK
excitations levels contributing to scattering processes
before violating unitarity. So therefore $\Lambda$ cannot be much bigger than
$1/R$.

  With the choice $\Lambda R = 20$ used in Ref. (\refcite{CMS}), 
  the spectrum of first
levek KK excitations is as follows. The heaviest particles are the $g^*$'s
which get a positive mass correction of about 30\% compared to the
tree level mass $1/R$ (we denote the KK excitation of a SM particle by a $^*$
superscript). Note that this correction is due almost entirely to
boundary terms; the bulk terms are very small, and actually negative.
The next to heaviest particles are the KK excitations of the SM quarks, which
get a mass correction of about 20\%. (There is a small mass splitting
between the tower associated with the left-handed fields $q^*$ and the 
tower associated with the right-handed fields $q'^*$ 
due to different electroweak interactions). 
Following are the heavy bosons excitations $W^*$ and $Z^*$, with
an 8\% mass correction, and the KK excitations of the leptons and neutrinos,
with a mass correction of below 5\%. The lightest KK particle (LKP) at the first
level will be the excitation of the photon $\gamma^*$, with mass very close
to the tree level value $1/R$.

This hierarchy has interesting implications. KK excitations of quarks and
gluons produced at hadron colliders will decay to the LKP $\gamma^*$
radiating Standard Model particles. The $\gamma^*$ will be stable, and it
does not interact with ordinary matter\footnote{One might think it possible
that $\gamma^*$ will interact with SM matter at higher orders. However, 
a remaining 
discrete $Z_2$ symmetry corresponding to  invariance of the Lagrangian under
reflections $y \leftrightarrow -y$ will still forbid processes where the KK 
conservation rule  is breaked by an odd number. This means one cannot produce
a single first level KK excitation at colliders, and also that
the LKP will be stable.}; it will therefore show up as missing
energy in the calorimeter. The signal for such processes will then
be several soft leptons or jets (coming from the decay of $q^*$'s and $g^*$'s)
plus large missing energy. We will discuss this in the next section.

As mentioned above, the appearance of boundary terms (terms in the Lagragian
proportional to $\del(y) + \del(\pi R -y)$) is a consequence of breaking
of translational invariance in the fifth dimension at the orbifold-fixed
points. Translation invariance implies momentum conservation in the 5th 
dimension; in the effective theory, this appears as KK number conservation. It
is not surprising then that at higher perturbation order, there will appear
terms in the Lagrangian which violate KK number conservation. 
 In the one-loop 5D effective Lagrangian, such terms will look like
$$
{\cal L}_{\psi \psi A}^{(1)} \ \sim \ 
\frac{g^2}{16 \pi^2} \ln\left( \frac{\Lambda^2}{\mu^2}\right)
\frac{\del(y) + \del(\pi R -y)}{2} \ g \bar{\psi} \gamma^\mu \psi A_\mu
$$
(these particular terms would come from correction to the 
fermion-fermion-gauge  boson vertices). By expanding in KK modes and 
integrating over the 5th coordinate, one can indeed see that the 
delta functions in the above expression causes the appearance of 
vertices where KK number conservation does not hold.
However,  the theory still has invariance
under reflections $y \leftrightarrow -y$; as a consequence, one is left
with KK parity as a conserved quantity (similar to R-Parity
in supersymmetric theories). This raises the interesting
possibility of a single KK level
two excitation being produced at colliders (since the mass of such
excitation is roughly the mass of two first level KK excitations, the 
phase space requirements are similar in both cases). Such a production
mode has potentially
important implications for our capability of making sure that the massive
particles one might find at colliders are part of  KK tower, thus showing
that the underlying theory is an extra-dimesional one.


\subsubsection{Stable LKP phenomenology}

As mentioned in the previous section, first level KK excitations of quarks and
gluons produced at a hadron collider will undergo a chain decay to the LKP,
radiating SM particles in the process. In order to study the phenomenology of
such events, one has to first derive the relevant decay modes and branching
rations for the heavy particles under consideration.

Details about the evaluation of these branching ratios can be found in Refs.
(\refcite{CMSexp,CMS}). 
Here we will summarize the results. The gluon excitations $g^*$
will decay to a quark pair (one quark being a SM particle, the other one being
a first levek KK excitation). The quark excitations (either produced directly
or through the decay of a $g^*$) will decay through electroweak interactions 
as follows. The singlet quark excitation will decay directly to the LKP:
$q'^* \rightarrow q \gamma^*$, since it does not couple directly to the 
SU(2) bosons, and it turns out that the mixing between the neutral SU(2)
boson field ($W^{0*}$)and the hypercharge field $B^*$ is very small (in other
words, the $\gamma^*$ is almost pure $B^*$).  The decay of the SU(2) doublet 
quark excitations will proceed mostly through the following channels:
\bea{chain} 
q^*\ & \rightarrow\  & q\ Z^*\ \rightarrow 
q\ \bar{l} \ l^*\ , \  q\ \bar{\nu} \ \nu^*\
\nonumber \\
 q^*\ & \rightarrow\ & q\ W^*\ \rightarrow 
q\ \bar{\nu} \ l^* \ , \ q\ \bar{l} \ \nu^* \ 
\eea
(and the charge conjugate ones),
with the branching ratio for the first decay chain being about 33\% and for the 
second one being about 65\%. Since the $Z^*$ is almost purely $W^{0*}$, the
branching rations for the $Z^*$ decay to neutrinos versus leptons are roughly
equal (also, it will decay predominantly to the left handed fermions
rather than to the right-handed ones). 
Furthermore, the KK excitations of leptons and neutrinos will decay
to the LKP: $l^* \rightarrow l \gamma^* ,\  \nu^* \rightarrow \nu \gamma^*$,
since this is the only decay channel kinematically allowed. 
Finally, a small fraction
of the $q^*$ excitations (about 2\%) decay directly to the LKP: 
$q^* \rightarrow q \gamma^*$.

The main experimental signal for the production and decay of KK excitations at
hadron colliders will then be the observation of events with multiple leptons
and jets 
of moderately high energies. Note that the leptons (which come from the decays
of the $W^*$ and $Z^*$ bosons) will have a maximum energy equal to the mass
difference between the $W^*$ and $\gamma^*$, or around 10\% of $1/R$. They will
not therefore be very hard, but they will have enough energy to pass the 
transverse momentum cuts 
used in experiments (which typically are in  the range of tens of GeV), and
there can be up to four of them in a single event. The jets produced
by the decays of quark and gluon excitations  will have energy similar to
that of leptons, but
due to the large backgrounds at hadron colliders, it is unlikely they
will be useful as signals for such events.  Finally, although the total
missing energy in the event (the energy carried away by the escaping
$\gamma^*$) will be large, the measurable missing energy (that is, the
transverse part) will also be of the order of the transverse momentum of
the observable leptons and/or jets. 

A preliminary analysis of collider signals for such models has been performed
in Ref. (\refcite{CMSexp}). Using the `gold-plated' decay mode with four leptons
in the final state (so called because the background for this signal is small),
one finds that at the Tevatron Run II the discovery reach is around
300 GeV (for an integrated luminosity of 10 fb$^{-1}$). At the LHC, with
100 fb$^{-1}$, one could discover Universal Extra Dimensions in this mode
if the masses of the first level KK excitations are less than about 1.5 TeV. 
Also, with new data coming from the Tevatron Run II, some analyses have been 
also performed for a signal with $\ge 3$ leptons in the final 
state\cite{smwang}. These
newer searches set a limit of $1/R \gtrsim 280$ GeV. 

We should make clear that any analysis we can perform is somewhat model
dependent. For example, the results mentioned above hold for the case 
$\Lambda R = 20$. If the cutoff $\Lambda$ is smaller, the mass splitting
between the first level KK excitations will correspondingly be smaller, and
therefore the energy of the final state leptons will decrease,
which makes it harder to differentiate the signal from background.
 The dependence on $\Lambda$ is logarithmic, therefore one might
expect that this will not be a big effect; however, there are 
constraints coming from unitarity which seem to indicate that the parameter
$\Lambda R$ should be as small as  5.
In this case the mass splittings will be only half compared
to the scenario discussed above,
which can potentially affect the ability to find a signal quite severely.

An even larger impact on the phenomenology of the model would result from
modifying some of the
renormalization  conditions on the 
boundary mass correction terms. For example, 
one might assume that values for the fermion $\del m$ parameters 
at scale $\Lambda$ differ by a finite (small) amount  from the values
of the boson $\del m$ parameters (in the Minimal UED model of 
Refs. (\refcite{CMS,CMSexp}) these mass correction terms are zero). Then, 
for example, the KK excitations of leptons might be heavier than the $W_1, Z_1$
boson excitations. The widths for the previously discussed 
decay modes of $Z^*, W^*$ will change; for example, the width for the decay
mode $Z^* \rightarrow \bar{l} l^* \rightarrow \bar{l} l \gamma^*$ will be
suppressed by the requirement that the intermediate $l^*$ fermion is off shell.
New competing decay modes might become important, like, for example
$W^* \rightarrow W \gamma^* \rightarrow l \nu \gamma^*$, which normally is 
suppressed by the small $W^{0*} - B^*$ mixing angle. A detailed analysis of
these possibilities has yet to be performed.

Another important question for the phenomenology of such models
is if one can determine the spins of the heavy particles being produced.
The leptons and jets + missing energy signal also arises in supersymmetric
models (see, for example, Refs. (\refcite{sugra-tev,sugra-lhc})), 
where the SUSY partners of the SM particles play the role of the 
first level KK excitations. For example, at hadron coliders one would
produce mostly  squarks and gluinos, which then can decay through steps
similar to those described for UED models (with neutralinos instead 
of the $Z^*$ and charginos playing the role of the $W^*$ bosons) to the 
lightest supersymmetric particle (the LSP), which is usually a neutralino.
Of course, the details of the decays and the experimental signal will depend
on the mass spectrum of the SUSY particles, which typically has large splitting
between the masses of the strong-interacting particles (gluinos, squarks)
and the weakly interacting ones.
However, for a particular class of SUSY models, one can have
a quasi-degenerate mass spectrum  which is similar to that of UED models, and
would lead to like signals.

Methods have been developed for measuring the spin of massive particles
for SUSY-like models\cite{Barr}. 
Such methods are based on the analysis of angular correlations
and invariant mass distributions
for the  visible lepton and/or jets, and they work reasonably well for mass
splittings in the range typically associated with SUSY. However, analyses
for the quasi-degenerate mass spectrum case which might arise from the UED 
model\cite{matchev05,Webber}
show that, due to the small energy of the observable leptons/jets, one cannot
easily distinguish between the SUSY and UED scenario 
with the same mass spectrum.

Finally, let us shortly discuss production of KK excitations of SM matter at 
$e^+ e^-$ colliders. For such production to take place, one would need the CM 
collider energy to be bigger than roughly two times the mass of the first KK
excitation (since conservation of KK parity requires either two 
level-one excitations or a level-two excitation in the final state). In
addition, one would look at production of KK excitations of
weakly interacting particles (like electrons or muons) rather than
strongly interacting ones. Note also that,
since the maximum CM energy attainable
at the next-generation linear colliders will probably
be below 3 TeV, this means
that if KK excitations are accesible to such a machine, it will be possible
to produce them at the LHC, too. A linear collider is therefore not a discovery
machine; however, due to the cleanliness of the environment in $e^+ e^-$ 
collisions, it will be possible to measure accurately the properties of the
particles under consideration (like mass, couplings and spin). For example,
the analysis in Refs. (\refcite{MatchevKong,Bhattacharyya}) 
indicates that it is possible
to cleanly differentiate between SUSY and UED signatures by using the angular
distributions of the decay products. 

\section{Matter and gravity in extra dimensions}

As we saw in the previous sections, experimental contraints require
the size for the extra dimensions in
which matter propagates to be order inverse TeV (or smaller). However,
if one wants to make use of extra dimensions to solve the hierarchy
problem (explaining why the Plank scale is so large in comparision to the
electroweak scale), one needs the size of the extra dimensions in which
gravity propagates to be as large as inverse MeV, up to inverse eV size
(depending on the number of dimensions of the space). 

One can reconcile these different requirements a number of ways. For example,
one might assume that the space is asymmetric: with one (or more) small extra
dimensions of size inverse TeV, in which both matter and gravity propagates,
and five (or less) larger extra dimensions  of size inverse eV, in which
only gravity propagates\cite{asym}.
While this scenario satisfies both requirements 
mentioned above, it does not bring anything new in terms of phenomenology.
KK number conservation rules still hold, and while first level excitations
of quark and gluons can decay to a first level graviton and a SM particle, the
strength for the coupling of individual gravitons is sufficiently small
that this decay channel will be highly suppressed. The resulting 
phenomenology arising from the production of KK excitations of matter
will be the same as discussed in the previous section; and the phenomenology
associated with the production of gravitons will be that of the standard
ADD model (see, for example, Refs. (\refcite{peskin-add,hewett-add,HLZ,GRW})).

A more interesting scenario would be a symmetric one in which 
all extra dimesions are large (order inverse eV); gravity propagates all
the way in this space (the bulk),
 the matter
fields, however, are restricted to a small region in the fifth dimension. 
This would be an extension of the ADD
model, where matter is confined on a 4D brane, with zero width in the extra
dimensions. In the scenario we shall discuss below, we take this brane
to have a finite width (of order inverse TeV) in the fifth dimension
(hence the name fat brane\cite{rigolin}). Such scenario could arise naturally
in a context in which one has some interaction which confines the matter
fields close to the 4D SM brane; if the strength scale of this interaction
is $M$, one would expect the wave function of matter fields to have 
a spread of order $1/M$ in the extra dimensions. However, we will not make 
any realistic attempt to try to model such interaction in our analysis, instead
assuming that the matter fields are confined in an infinite potential well
of width $1/M$ centered on the SM brane. 


Such scenario has interesting phenomenological 
consequences\cite{Rizzo-ued,cmn,cmn2,marius1,Marius2}.
Since one restricts matter on the brane, momentum
conservation in the fifth dimension does not hold anymore (the extra
momentum associated with the radiation of a graviton in the bulk can 
be picked up by the brane). Then, KK number conservation does not
hold anymore for matter-gravity interactions, and the first level KK 
excitations of matter can decay by radiating gravitons. Since the mass
splitting in the graviton tower is quite small (inverse eV), the resulting
decay widths can be quite large, resulting in gravitational decay modes
competing with the decay channels
induced by first level mass splittings.
Moreover, one can have production of a single KK matter excitation, mediated
by the gravity interactions (with either virtual gravitons as intermediate
states, or with real gravitons in the final state). The analysis of such 
possibilities will be the topic of the following sections.

\subsection{The gravity sector}

We will start by reviewing the reduction of gravitation from a $4+N$
dimensions theory to an effective 4D theory (in other words, expanding
the gravitational fields in KK modes, similar to the procedure used
for the matter fields in section {\bf 2.1}).
We will follow here mostly the analysis
in Ref. (\refcite{HLZ}). 
The extra dimensions are compactified on a torus $T_N$, with
the linear length $r$ (hence the compactification radius will be $r/2\pi$).
The graviton field (associated with the linearized metric fluctuations) 
will then have a KK expansion
\beq{grav_exp} \hh_{\hmu\hnu}(x,y)\ =\ V_N^{-1/2} \sum_{\vec n}
\hh_{\hmu\hnu}^{\vec{n}}(x)\ \exp\left(i \frac{2\pi
\vec{n}\cdot\vec{y}}{ r}\right)\ , \eeq 
where $V_N= r^N$ is the volume of the $N$-dimensional torus.
The 'hat' denotes quantities which
live in 4+$N$ dimensions: $\hmu, \hnu = 0,\ldots,3,5,\ldots 4+N$,
while $\mu, \nu = 0,\ldots,3$. $\vn$ is the KK vector associated
with a particular graviton excitation $\vn = \{n_i\},\ i = 5,\ldots 4+N$.   
At each KK level, the graviton field is
decomposed into 4D tensor, vector and scalar components by:

\beq{hh_def} \hh_{\hmu\hnu}^\vn \ =\ \left(\begin{array}{cc}
h_{\mu\nu}^\vn+\eta_{\mu\nu}\phi^\vn & A_{\mu i}^\vn \\
A_{\nu j}^\vn   &  2 \phi_{ij}^\vn
\end{array}\right)\ .
\eeq

Upon writing the $4+N$ Pauli-Fierz gravitational lagrangian and expanding
in KK modes, one sees that not all the $h_{\mu\nu}^\vn,\ 
A_{\mu i}^\vn,\ \phi_{ij}^\vn$ field components are
independent. In order to eliminate the unphysical degrees of freedom (d.o.f.),
one has to get rid of the  lagrangian
invariance to general coordinate transformations
by imposing some constraints  (equivalent to picking a particular gauge).  
Thus, we use the de Donder condition
\beq{donder}
\partial^\hmu (\hh_{\hmu\hnu}-\frac{1}{2}\eta_{\hmu\hnu}\hh) =0 \ ,
\eeq
(with $ \hh = \hh^\hmu_{\ \hmu}$), plus
\beq{gauge_fix}
n_i A_{\mu i}^\vn = 0,  \ \ n_i \phi_{ij}^\vn = 0 \ ,
\eeq
at each KK level. These constraints will eliminate $2 D$ spurious
d.o.f. (where $D= 4+N$)
out of the $D(D+1)/2$ appearing in $\hh_{\hmu\hnu}^\vn$, leaving
$D(D-3)/2$ physical degrees of freedom at level $\vn$, associated 
with a massive spin 2 graviton (5 d.o.f.), $N-1$ massive vector
bosons ($3\times(N-1)$ d.o.f.) plus $N(N-1)/2$ massive scalars
with 1 d.o.f. each. The physical gravity fields are denoted by tilde,
and are given by:
\bea{h_def}
\tih_{\mu \nu}^\vn & = & h_{\mu \nu}^\vn -
\omega \left( \frac{ \del_\mu \del_\nu}{m_\vn^2} - {1\over 2}
\eta_{\mu \nu}  \right)  \tp^\vn \nonumber \\
\tA_{\mu i}^\vn & = & A_{\mu i}^\vn \nonumber \\
{1 \over \sqrt{2}} \tp_{ij}^\vn  & = & \phi_{ij}^\vn +
{ 3 \omega a\over 2 }
 \left( \delta_{ij}- { n_i n_j \over \vn^2} \right) \tp^\vn
\eea
and $ \tp^\vn = (2/ 3\omega ) \phi^\vn $.
Here $\phi^\vn = \phi^\vn_{ii}$ (
same for the tilde fields),
$\omega = \sqrt{2/3 (N+2)}$, and
$a$ is a solution of the equation $3(N-1)a^2 + 6 a = 1$ (as shown in
Ref. (\refcite{HLZ})).
The physical fields also satisfy
\beq{pol_sum}
n_i \tA_{\mu i}^\vn = 0,  \ \ n_i \tp_{ij}^\vn = 0 \ ,
\eeq
which are the equivalents of the gauge invariance relation 
$k_\mu A^\mu(k) = 0 $ for 4D gauge fields.

\subsection{Gravity - matter interactions}

Once the physical excitations of the graviton fields are known, 
one can derive  the matter-gravity interactions. 
For the case of mattes restricted on the 4D brane (the ADD scenario) 
the interaction lagrangian has been derived in Refs. (\refcite{HLZ,GRW}).
We review
here the case when matter propagates in 5 dimensions, following the
analysis in Ref. (\refcite{Mitov}) (interaction rules for the case
when gravity and matter propagate on the same space have also 
been derived in Ref. (\refcite{Feng-gr})).

Since gravity couples to
matter through the energy-momentum (EM) tensor, 
the $D$ dimensional action will be
\beq{act_gen}
{\cal S}_{int} = -{{\hat \kappa} \over 2}
\int d^D x \ \delta(x^6) \ldots \delta(x^N)\
 \hh^{\hmu\hnu} T_{\hmu\hnu} \ ,
\eeq where $\hat{k}$ is the $D$ dimensional gravitational coupling constant,
and
\beq{em_tensor}
T_{\hmu\hnu} \ = \ \left( -\hat{\eta}_{\hmu \hnu} + 2 \frac{\del 
{\cal L}_m}{\del \hat{g}^{\hmu \hnu} }\right)_{\hat{g} = \hat{\eta}} \ ,
\eeq
${\cal L}_m$ being the matter lagrangian.
Expanding the gravity field in KK modes and writing the
matter EM tensor in terms of its $(\mu \nu),\ (\mu 5)$ and $(55)$
components, we obtain
$$
 {\cal S}_{int} = -{\kappa \over 2} \int d^4 x
\int_0^{\pi R} dy  \sum_{\vn} \left[ \left( h^\vn_{ \mu \nu} +
\eta_{\mu \nu} \phi^\vn \right) T^{\mu \nu} -2 A^\vn_{\mu 5}
T^\mu_5 + 2 \phi^\vn_{55} T_{55} \right] e^ {2\pi i { n_5 y\over
r}}. \nonumber 
$$
Here $\kappa$ is the four-dimensional Newton's constant 
$\kappa \ \equiv \ \sqrt{16 \pi G_N} \ = \ V_n^{-1/2} \hat{\kappa}$.

In terms of the physical gravity fields, the effective 4D
Lagrangian is then: 
\bea{l_4d} {\cal L}_{int} & = & -{\kappa \over 2}
\sum_{\vn} \int_0^{\pi R} dy \ \left\{ \left[ \tih^\vn_{ \mu \nu}
+ \omega \left( \eta_{\mu \nu} + \frac{\der_\mu \der_\nu}{m_\vn^2}
\right) \tp^\vn \right]
T^{\mu \nu}   - \right.  \nonumber \\
& & \left. 2 \tA^\vn_{\mu 5} T^\mu_5 \ + \
 \left(  \sqrt{2} \tp^\vn_{55} - \xi \tp^\vn \right) T_{55}
\right\} e^ {2\pi i { n_5 y\over r}} \ ,
\eea
where $\xi = 3 \omega a  ( 1- n_5^2 / \vn^2)$. The EM  tensor
for different types of matter can be evaluated using (\ref{em_tensor});
for example, for a scalar field $\Phi$ one would obtain:
\beq{t_scal}
T^{ S}_{MN}\ =\ -\eta_{MN} ( D^R\Phi^\dagger D_R\Phi
        - m^2_\Phi\Phi^\dagger\Phi )
    +D_M\Phi^\dagger D_N\Phi
    +D_N\Phi^\dagger D_M\Phi\ ,
\eeq where $M,N,R$ are indices which run from 0 to 5, and $D_M =
\der_M + i g B^a_M T^a$ is the covariant derivative. It is convenient to
define the projections of the matter EM tensor on the $\vn$-th graviton state by
$$ T_{MN}^{n_5} (x) =
\int_0^{\pi R} dy\ T_{MN}(x,y)\ e^ {2\pi i { n_5 y\over r}} \ . $$ 
The interaction lagrangian would then be
\bea{l_4d2}
{\cal L}_{int} & = &  -{\kappa \over 2} \sum_{\vn}
\left\{ \left[ \tih^\vn_{ \mu \nu} +
\omega \left( \eta_{\mu \nu} + \frac{\der_\mu \der_\nu}{m_\vn^2} \right)
\tp^\vn \right]
T_{n_5}^{\mu \nu}   - \right.  \nonumber \\
& & \left. 2 \tA^\vn_{\mu 5} 
T_{n_5 5}^{~\mu}  +
 \left(  \sqrt{2} \tp^\vn_{55} - \xi \tp^\vn \right) T^{n_5}_{55}
\right\}. \eea

One would then proceed to expand the matter fields in $T_{MN}$ in KK modes,
and work out the Feynman interaction rules. The resulting expressions
are presented in Ref. (\refcite{Mitov}),
and, since they are quite complicated, we do not
show them here. We just note that the coupling for 
each vertex gets multiplied by a form factor
\beq{ff} {\cal F}^{(\hbox{f}_1,
\hbox{f}_2...)}_{l_1,l_2...|n}  \sim
 \int_0^{\pi R}
 dy \  \hbox{f}_1 \biggl( {l_1 y\over R}\biggr) \
 \hbox{f}_2 \biggl( {l_2 y\over R}\biggr) \ (\ldots) \
 \exp\biggl( 2 \pi i{n y \over r}\biggr) ,
\eeq
where the functions $\hbox{f}_i()$ can be sin() or cos() (there
may be more that 2 such functions at a vertex, depending on how
many matter fields participate in the interaction). These form factors 
describe
the superposition of the wave functions of the interacting particles in
the fifth dimension. We would like to point out that the information
about the profile of the wave function (thus about the mechanism
which keeps matter stuck on the brane) is encoded solely 
in these form factors; the results obtained in Ref. (\refcite{Mitov})
for the gravity matter interactions can be applied directly for cases 
in which the confinig potential had different forms, one needs just
to derive the respective eigenfunctions (which will not be simply
sine or cosine) and recompute the form factors. In the limiting case
when $\pi R = r$, one obtains the interactions for the case when gravity 
and matter propagate on the same space\cite{Feng-gr},
and where many of these form-factors
will be zero due to KK number conservation. 
 
\subsection{Gravity-mediated decays of KK particles}

One can use the interaction rules derived in the previous section to 
evaluate the decay widths of matter KK excitations to gravitons and
Standard Model matter. One would then obtain\cite{Mitov}: for the decay
of a KK fermion to a single graviton\footnote{We correct here some typos
appearing in Ref. (\refcite{Mitov}). The form-factor appearing 
for decays mediated by the vector gravitons is 
 ${\cal F}^s_{l|n}$  rather than  ${\cal F}^c_{l|n}$.}
\bea{fer_wid}
\Gamma (q^l \rightarrow q h^\vn) & = & | {\cal F}^c_{l|n} |^2 \
 \frac{\kappa^2}{2\times 384 \pi} \frac{M^3}{x^4}
\left[ \left( 1 - x^2 \right)^4 \left( 2 + 3 x^2 \right)
\right ] \nonumber \\
\Gamma (q^l \rightarrow q A^\vn) & = & | {\cal F}^s_{l|n} |^2 \
 \frac{\kappa^2}{2\times 256 \pi} M^3
\left[ \left( 1 - x^2 \right)^2 \left( 2 +  x^2 \right)
\right ] \times P_{55} \nonumber \\
 \Gamma (q^l \rightarrow q \phi^\vn) & = & | {\cal F}^c_{l|n} |^2 \
 \frac{\kappa^2}{2\times 256 \pi} M^3 \left( 1 - x^2 \right)^2
\left[ c_{11} \frac{( 1 - x^2 )^2}{x^4} \right.
\nonumber \\
& & \left. + 2 c_{12} \frac{ 1 - x^2 }{x^2} + c_{22} \right] \ .
\eea
Here $M$ is the mass of the matter KK particle $M=l/R$, $m_\vn$ is
the mass of the graviton, and $x = m_\vn/M$. The coefficients
$P_{55}$ and $c_{ij}$ appear because, as described
in section 3.1, not all $\tA_i, \tp_{ij}$ fields
are independent. To eliminate the spurious
degrees of freedom one uses
the (extra dimensional) polarization vector $e^k_i$ and tensor
$e^s_{ij}$ as in Ref. (\refcite{HLZ}), with:
\bea{pol_sums}
P_{55} & = & \sum_{k=1}^{N-1} e_i^k e_j^{k*} \ \delta_{i5} \delta_{j5}
 = 1-\frac{n_5^2}{\vn^2} \nonumber \\
c_{11} & = &  \sum_{s=1}^{N(N-1)/2} e_{ij}^s e_{kl}^{s*} \
\omega^2 \delta_{ij} \delta_{kl} = \omega^2 (N-1) \nonumber \\
c_{12} & = &  \sum_{s=1}^{N(N-1)/2} e_{ij}^s e_{kl}^{s*} \
\omega \delta_{ij}
(\sqrt{2} \delta_{k5} \delta_{l5} - \xi \delta_{kl} ) =
-\frac{2}{N+2} P_{55} \nonumber \\
c_{22} & = &  \sum_{s=1}^{N(N-1)/2} e_{ij}^s e_{kl}^{s*} \
(\sqrt{2} \delta_{i5} \delta_{j5} - \xi \delta_{ij} )
(\sqrt{2} \delta_{k5} \delta_{l5} - \xi \delta_{kl} ) =
\frac{2(N+1)}{N+2} P_{55}^2 \ .
\eea
For the decay of a KK gauge boson excitation, the following
results are obtained: 
\bea{bos_wid} \Gamma (B^l \rightarrow B
h^\vn) & = & | {\cal F}^c_{l|n} |^2 \
 \frac{\kappa^2}{3\times 96 \pi} \frac{M^3}{x^4}
\left[ \left( 1 - x^2 \right)^3 \left( 1 + 3 x^2 + 6 x^4 \right)
\right ] \nonumber \\
\Gamma (B^l \rightarrow B A^\vn) & = & | {\cal F}^s_{l|n} |^2 \
 \frac{\kappa^2}{3\times 32 \pi} {M^3 \over x^2}
\left[ \left( 1 - x^2 \right)^3 \left( 1 +  x^2 \right)
\right ] \times P_{55} \nonumber \\
 \Gamma (B^l \rightarrow B \phi^\vn) & = & | {\cal F}^c_{l|n} |^2 \
 \frac{\kappa^2}{3\times 128 \pi} M^3 \left( 1 - x^2 \right)^3
\left[ c_{11} \frac{1}{x^4} + 2 c_{12} \frac{ 1 }{x^2} + c_{22}
\right] \ . \eea 

The form-factors appearing in the above expressions are
\beq{dec_ff}
{\cal F}^{(c,s)}_{l|n} \ = \ \frac{\sqrt{2}}{\pi R} \int_0^{\pi R}
 dy \  (\cos, \sin) \biggl( {l y\over R}\biggr) \
 \exp\biggl( 2 \pi i{n y \over r}\biggr) \ 
\eeq
(for simplicity, $n$ stands for $n_5$ here), with
\beq{dec_ff1}
|{\cal F}^c_{l|n} |^2\ = \ 
\frac{4}{\pi^2}\frac{x_y^2}{(1-x_y^2)^2}\left[1+\cos(\pi
x_y)\right] \ , \ \
 |{\cal F}^s_{l|n} |^2\ = \ \frac{|{\cal F}^c_{l|n} |^2}{x_y^2} \ ,
\eeq
where $x_y = 2 \pi n R/ (l r) = m_{\vn}(n/\sqrt{\vn^2})/M$.

An interesting observation about these decay widths is that, 
not taking into account the form factor, for small
graviton mass some of them behave like $1/m_g^4$. This is somewhat
surprising; based on naive dimensional analysis, one would expect the
decay widths for small graviton mass to behave like $M^3/M_{Pl}^2$
(the depencence on $M_{Pl}$ arises through the gravitational coupling 
constant: $\kappa^2 = 16\pi/M_{Pl}^2$). One instead gets an enhancement
factor $(M/m_g)^4$, which, for order eV graviton masses will be quite
substantial. 

One can trace the appearance of this factor to terms
proportional to $k^4/m_g^4$ in the polarization sum for a graviton 
of momentum $k$. Typically, such terms should be zero, on account
of momentum conservation  laws. For example, the amplitude for the decay
of a quark KK excitation to a spin-2 graviton can be written as\cite{Mitov}
$$ {\cal M}(q^l \rightarrow q h^\vn) \ \sim \ 
\langle q | T_{\mu \nu} | q^l \rangle \ \eps^{\mu \nu}(k) \ ,$$ 
where $T_{\mu \nu}$ is the EM tensor of the quark field (the 4D components), 
and $ \eps^{\mu \nu}(k)$ is the polarization vector of the graviton.
Hence
$$
\sum_{\hbox{spin}}|{\cal M}(q^l \rightarrow q h^\vn)|^2 \ \sim \ 
\langle q | T_{\mu \nu} | q^l \rangle \langle q | T_{\rho \sig} | q^l \rangle\ 
B^{\mu \nu , \rho \sig}(k)
$$
(the expression for the spin-2 polarization sum $B^{\mu \nu , \rho \sig}(k)$
can be found in Ref. \refcite{HLZ}). Now, terms $\sim k^\mu k^\nu/m_g^2$ in
$B$ should give zero contribution, since the energy-momentum conservation
in 4D
should insure that $k^\mu T_{\mu \nu} = 0$. However, our theory is 
five-dimensional, and the relation which holds is $k^M T_{M N} = 0$.
Since $k^5$ is the difference of the 5D components of the momenta of
 matter particles
(their KK masses), we can rewrite the above $k^\mu T_{\mu \nu} = 
- k^5 T_{5 \nu} \sim M$.
(One can also verify directly, for example by using the expression \ref{t_scal}, that $\der^\mu T^S_{\mu \nu}(x) \sim 1/R$). 
Hence, terms $\sim k^\mu k^\nu/m_g^2$ in the polarization sums
will give contributions proportional to $M^2/m_g^2$, thus enhancing the
gravitational decay width of KK matter excitations\footnote{The
mechanism is similar to the decay of the top in
the Standard Model. In that case, the breaking of electroweak symmetry
which gives mass to quarks implies $k^\mu {\cal M}_\mu \sim m_t$
(${\cal M}_\mu$ beig the decay amplitude),
rather than $k^\mu {\cal M}_\mu =0$, as required by gauge invariance.
As consequence,
 the top decay width is  $\sim \alpha m_t^3/M_W^2$, rather
than $\sim \alpha m_t$, as expected from naive dimensional analysis.}.
This enhancement 
is quite important for $N=2$ extra dimensions, and implies that
in this case the KK excitations decay mostly to 
light gravitons.


To compute the total gravitational decay width, one has to sum over 
all the gravitons with mass smaller that the decaying particle.
In the case of small splitting between graviton excitation masses,
such a sum can be replaced by an integral:
\beq{g_sum}
\sum_{\vn} \ \rightarrow \ \frac{1}{(m_{g}^{0})^N}
\int m^{N-1} d m \ d \Omega 
\ = \ \frac{M_{Pl}^2}{M_D^{N+2}} 
\int m^{N-1} d m \ d \Omega \ ,
\eeq 
where $m_g^0 = 2\pi/r$ is the mass splitting in the graviton tower (also
equal to the mass of the lightest graviton excitation), 
$M_{Pl}$ and $M_D$ are the 4D, respectively (4+N) dimensions Planck scales
and we have used the ADD relation
\beq{ADDf}
M_{Pl}^2 \ = \ M_D^{N+2} \left( \frac{r}{2\pi}\right)^N \ .
\eeq 
Note here that the definition of $M_D$ varies somewhat through the literature;
the choice 
$\bar{M}_{Pl}^2  =  M_D^{N+2} ({r}/{2\pi} )^N$
is also sometimes used\cite{GRW}, with the reduced Plank mass 
$\bar{M}_{Pl} = {M}_{Pl}/\sqrt{8\pi}$. We also leave the differential angular 
element $d \Omega$ in the equation (\ref{g_sum}), since in our model the
integrand is not invariant with respect to rotations in the extra-dimensional
coordinates (the $y$ direction is special).

As we saw above, decay through radiation of light gravitons is preffered
by the amplitude square. However, due to the geometry of the space, there
are many more heavy gravitons around (the number of gravitons with
masses in  a range $(m, m+\Delta m)$ is proportional to 
$m^{N-1}\Delta m/(m_g^0)^N$). To see which effect dominates, we have to look
at the overall  behavior of the integrand with respect to $m_g$.
\beq{DW_tot}
\Gamma_h \ = \ \frac{M_{Pl}^2}{M_D^{N+2}} 
\int \Gamma^\vn \ m^{N-1} d m \ d \Omega \  \sim \  
 \frac{M^5}{M_D^{N+2}} 
\int \frac{m^{N-1} m^2}{m^4} d m \ d \Omega \ ,
\eeq
for spin two and scalar gravitons, and small masses $m$ 
(the $m^2$ term in the denominator is due the behaviour
of the form factor,  which is 
$|{\cal F}^c_{l|n} (x)|^2 \sim x^2$, for $x \ll 1$). For vector gravitons
in the final state, the amplitude behaves as $(M/m_g)^2$, but the form-factor
$ |{\cal F}^c_{l|n} (x)|^2 $ is close to one.
Hence we see that 
for $N=2$ the small gravitons will dominate (due to the $1/m_g$ integrand),
while for large values of $N$ the heavy gravitons with $m \sim M $ 
will account for most of the total gravitational decay width.

\begin{figure}[t!] 
\centerline{
   \includegraphics[height=2.8in,width=3.in]{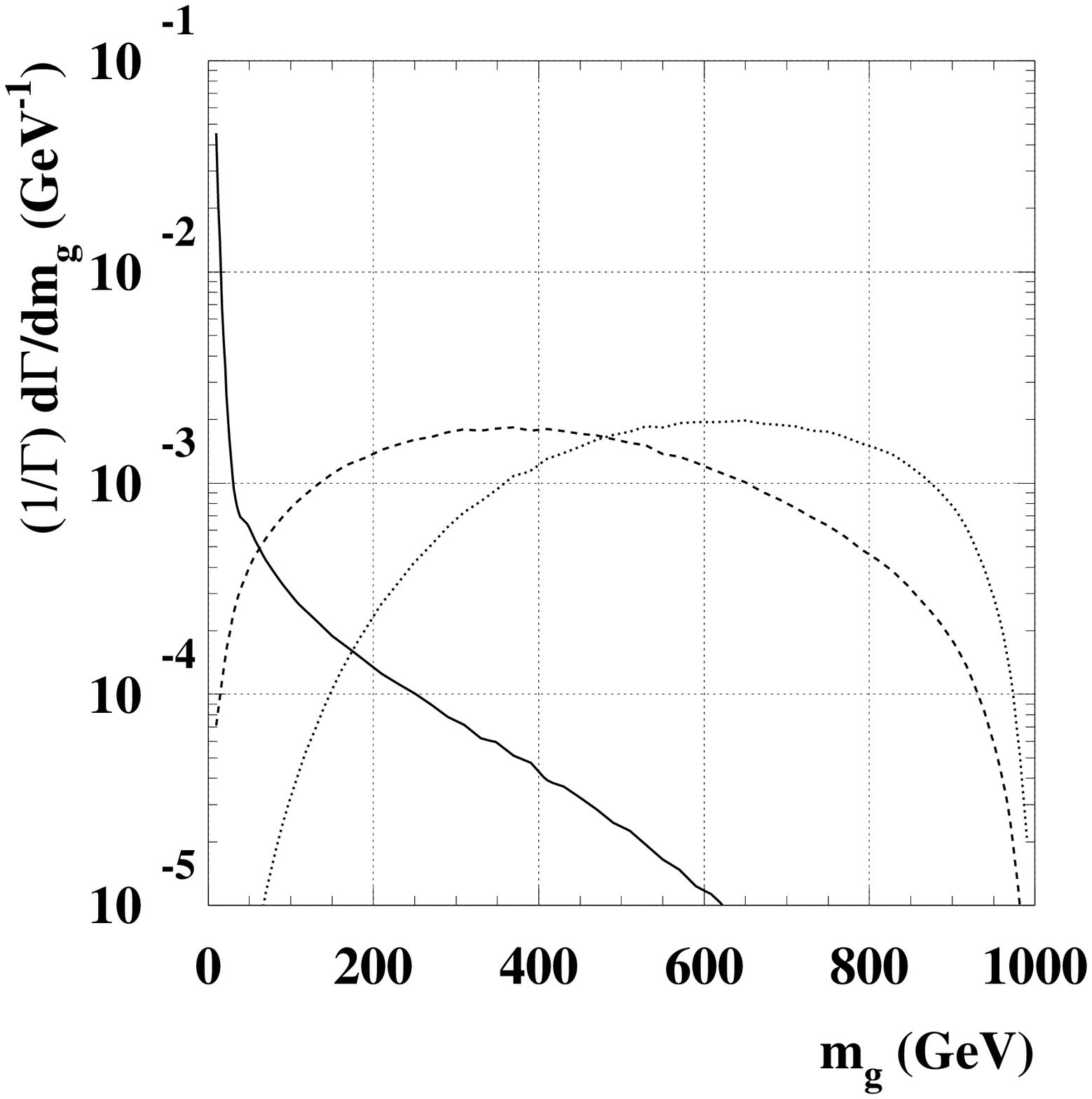}
   \includegraphics[height=2.8in,width=3.in]{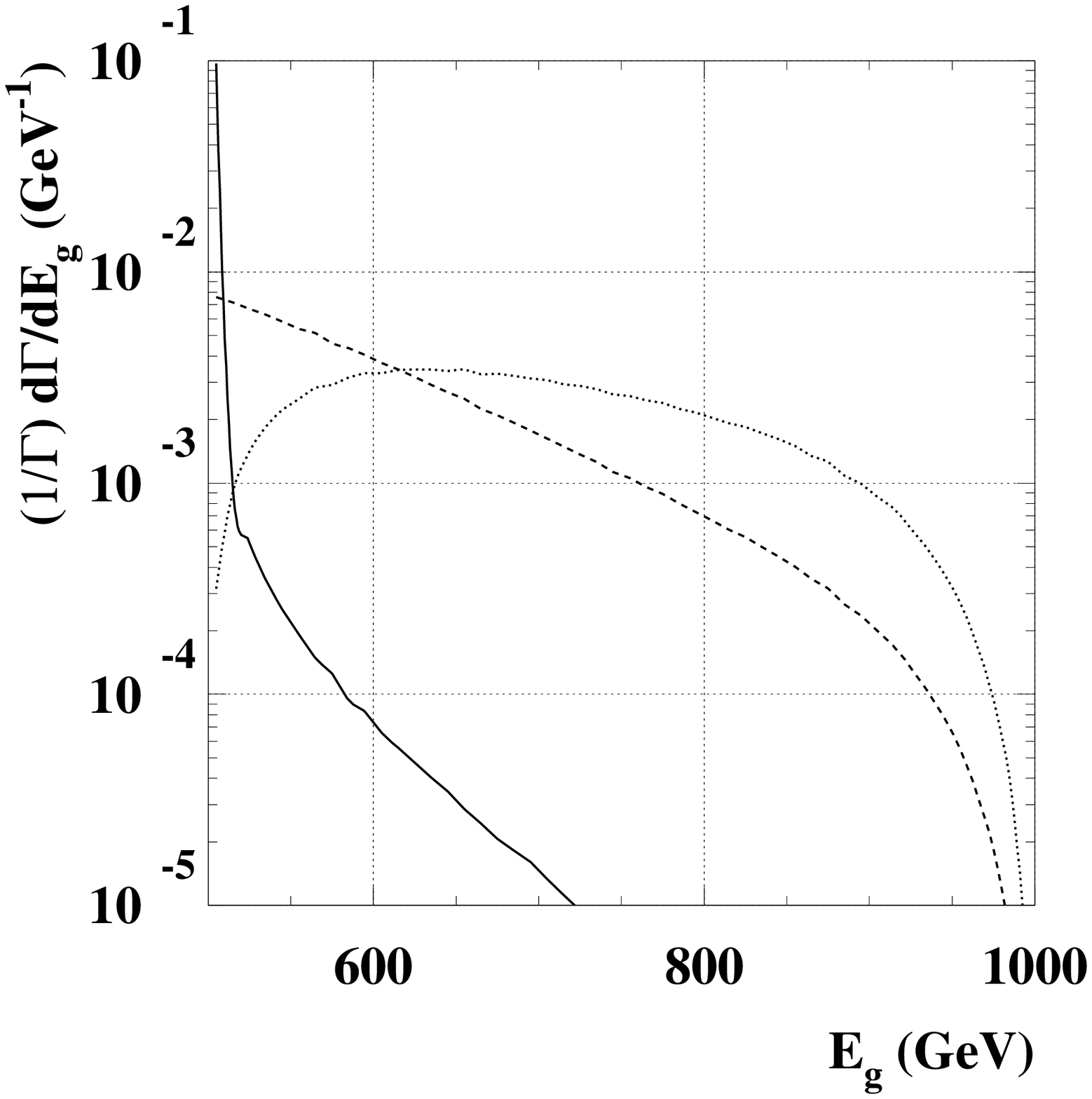}
   }
\caption{ Mass distribution (left) and energy distribution (right)
for the graviton radiated in the decay of one matter KK excitation
with mass 1 TeV. Straight lines corresponds to $N=2$ extra
dimensions, dashed lines to $N=4$, and dotted lines to $N=6$. 
The integral of the area under the individual curves is equal to 1.}
\label{decdist}
\end{figure}
	
Note
also that the form factor associated with the decays to spin-two and scalar
gravitons has a suppression effect on the cross-section, even
if the masses of the final state gravitons are large
(this happens because $n_5$ is
typically much smaller than $|\vn|$). This does not matter very much
for $N=2$, because of the $(M/m_g)^4$ enhancement  factor. In particular,
for $N=2$, since mostly the lightest gravitons contribute,
 one can estimate the relative decay widths to spin-two, vector
and scalar gravitons from Eqs. (\ref{fer_wid}), (\ref{bos_wid}). One
obtains
$\Gamma_h: \Gamma_A :\Gamma_\phi \sim 1:0:1/8$ for fermions, and
$\Gamma_h: \Gamma_A :\Gamma_\phi \sim 1:3:1/8$ for bosons (the smallness
of the decay width to vector gravitons for fermions is due to the
absence of the enhancement factor $(M/m_g)^2$).  
However,
for $N \ge 3$ this means that the decay width to spin-two and
scalar gravitons is suppressed in comparision to the decay width 
to vector gravitons.

As an illustration, we plot in Fig. \ref{decdist} (left panel)
the partial decay width as a function of graviton mass
(computed for the decay of a 1 TeV KK excitation of a fermion).
One sees that the above analysis is 
correct, and the total decay width for $N=2$ is due mostly to 
light gravitons, while
for $N=4, 6$ heavier gravitons dominate. In the right panel, we plot the 
energy of the final state graviton (which has implications for the 
collider phenomenology of the model, as it will be discussed in the next
section). Thus, for $N=2$, the graviton energy is
equal to about half the mass of the particle
(since for this purpose the graviton mass can be taken to be close to zero),
while for higher $N$, the graviton has typically an average energy closer to 
three quarters of the mass $M$ of the decaying particle.

\begin{figure}[t!] 
\centerline{
   \includegraphics[height=3.in]{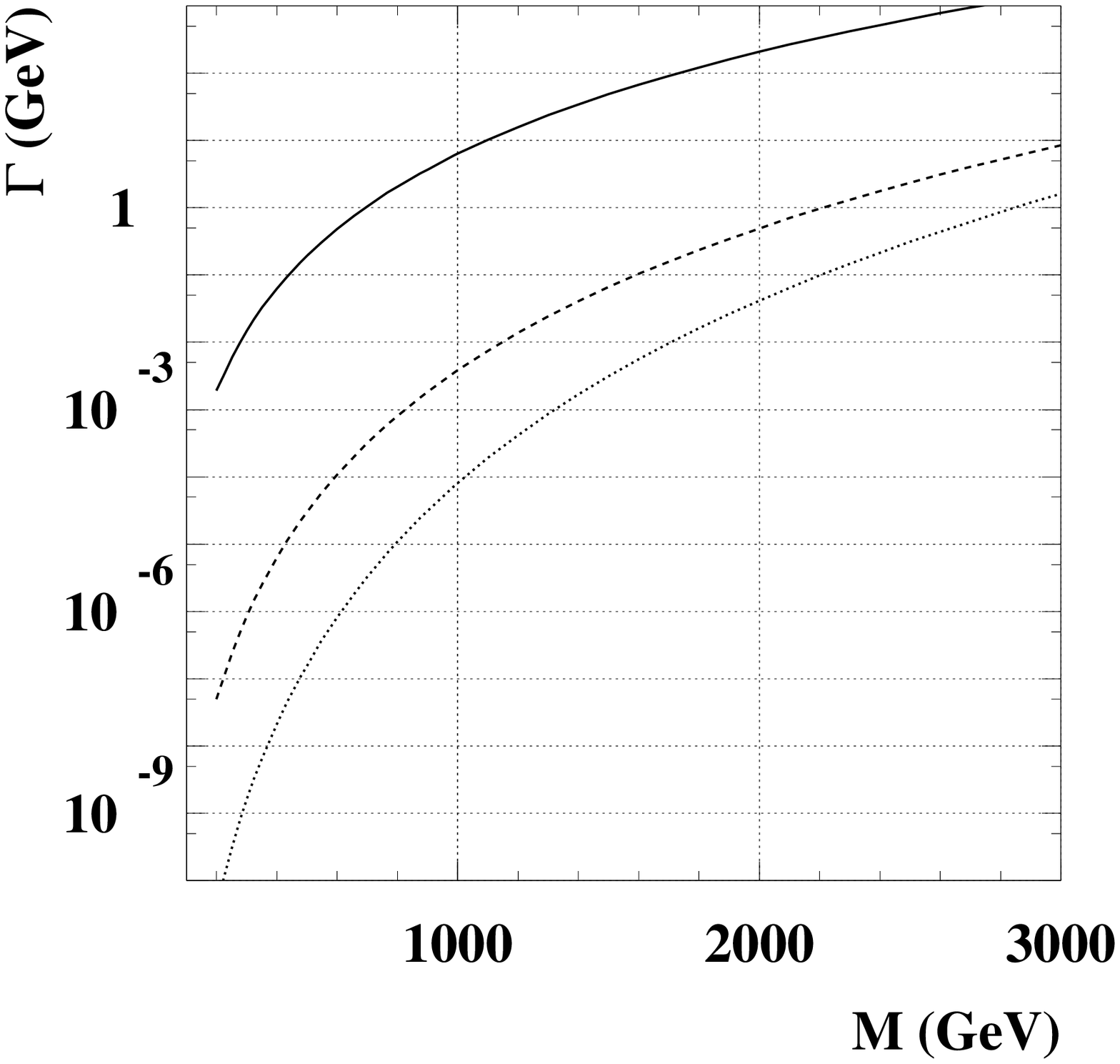}
   \includegraphics[height=3.in]{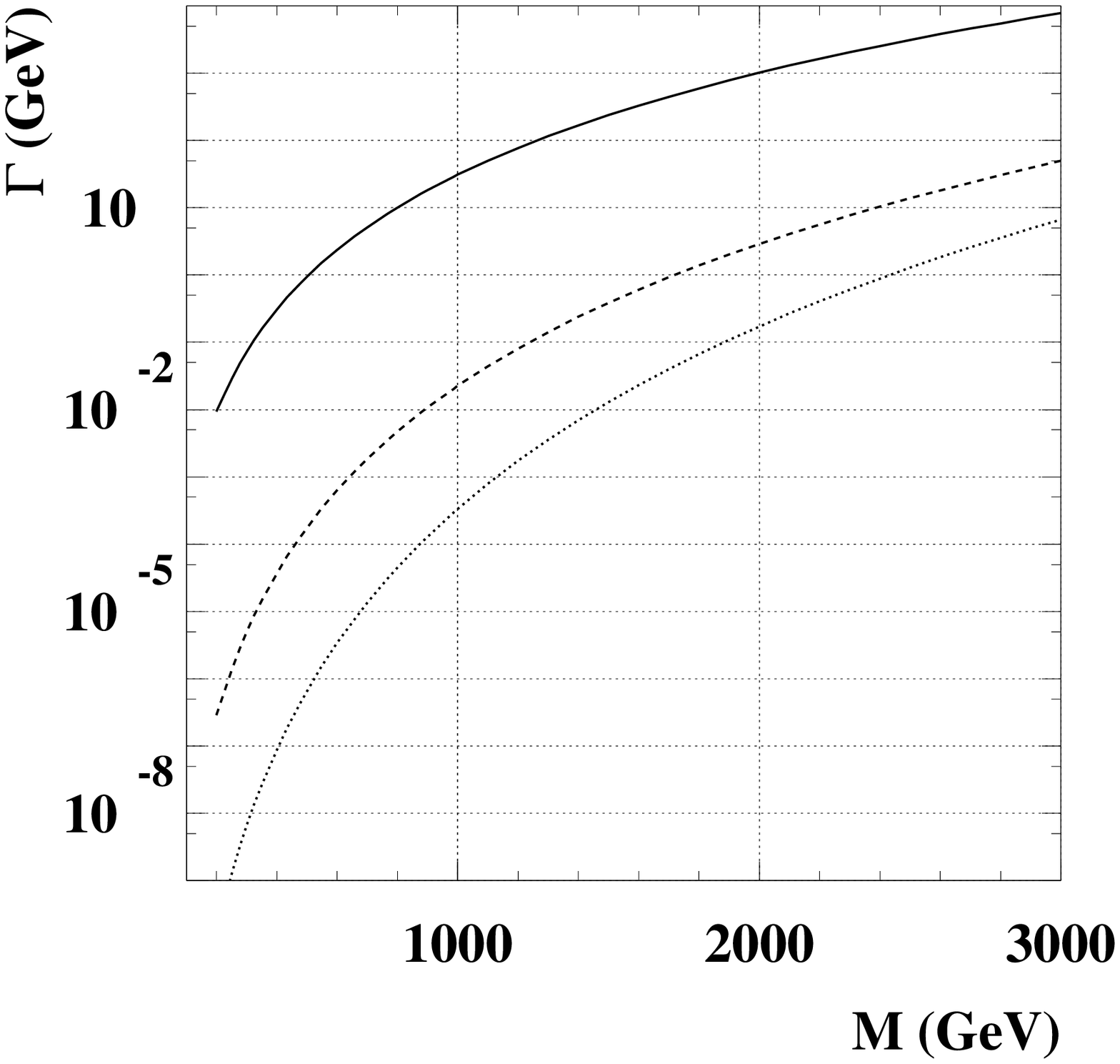}
   }
\caption{ Decay widths for KK fermions (left) and bosons (right)
as a function of the particle mass. Straight lines corresponds to $N=2$
extra dimensions, dashed lines to $N=4$, and dotted lines to $N=6$. Here 
$M_D$ is taken to be 5 TeV.}
\label{kkwid}
\end{figure}

In Fig. \ref{kkwid} we plot the total gravitational decay widths for 
KK excitations of fermions (left panel) and gauge bosons (right panels).
We take $M_D = 5$ TeV, and let $M$ to vary from 200 GeV up to 3 TeV. We
see that generally the decay widths are larger for $N=2$ (due to smaller
mass splitting between the graviton masses, as well as due to the enhancement
factor discussed above). Also, even for $N=6$, the decay widths are typically
large enough that the particles will decay into detector. Moreover, among
the first level KK excitations, the
gravitational decay widths are comparable with the decay widths due to mass
splittings. A direct comparision of the widths associated with these two types
of decays can be found in Ref. (\refcite{cmn2}); however, one must keep
in mind that the results for gravitational decays depends strongly on
the parameters $M, M_D$, and as such, care must be used when commenting
on the relative importance of the two decay modes 

In order to facilitate such a comparision,
 we shall give here on the dependence of the gravitational decay
widths on the parameters $M$ and $M_D$. For $N=3, \ldots , 6$,  
the behaviour of the dominant terms (large $m$) in the integral (\ref{DW_tot})
goes like $m^{N-1}M^3/M_D^{N+2}$. Since the upper limit for $m$ in the integral
is $M$, this would lead to a behaviour $\Gamma_h \sim M^{N+3}/M_D^{N+2}$
(consistent with the results shown in Fig. \ref{kkwid}).
On the other hand, for $N=2$ only the gravitons with lowest masses typically
give a nonnegligible contribution to the total width, and therefore $\Gamma_h$
scales like the decay width to the lowest mass graviton
$\Gamma_h \sim (1/M_{pl}^2)  M^5/(m_g^0)^2 \sim  M^5/M_D^4$ (where we have
made use of the ADD relation (\ref{ADDf}) ). These relations can 
be verified by numerical computations.

\subsection{Phenomenology of gravity-mediated decays}

In this section we will discuss the phenomenology of models in which
the decay of matter KK excitations can be mediated by gravity. In such  cases,
the experimental signal observed will be the SM particle(s) corresponding to 
the KK excitation(s) which decays gravitationally, and missing energy taken
away by the gravitons (which interact too weakly 
to be observed in the detector).
By contrast with the case where the LKP is stable,
the energies of the SM particles
observed (either quarks and gluons, which will appear as jets, or leptons and
photons) will be large, since they are the final products of the decay of
a massive particle (the KK excitation). One therefore obtains a strong signal 
in such scenarios, which makes it easy to observe the 
new particles and/or constrain the model.

\begin{figure}[b!] 
\centerline{
   \includegraphics[height=3.in]{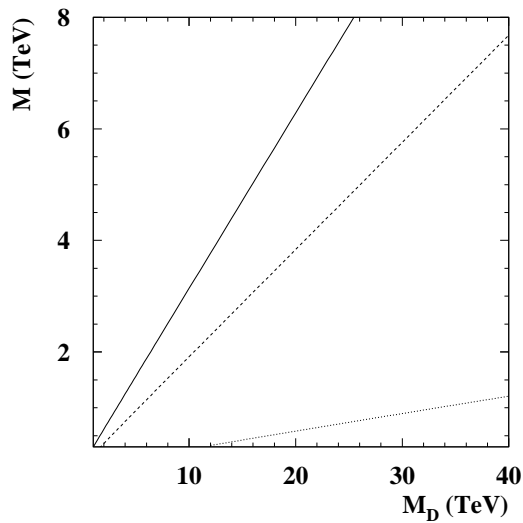}
   }
\caption{Regions in the parameter space where decays to $\gamma^*$
dominate versus the gravitational decays for $q^*, g^*$. 
The solid, dashed and
dotted  lines correspond to $N$ = 6, 4 and 2  extra dimensions.}
\label{split}
\end{figure}

Depending on the relative strength of the decay channels of the KK 
excitations, one can 
identify three separate scenarios for the phenomenological signal.
First is the case when the gravitational decays dominate. Then, KK
excitations of quarks and gluons decay to SM quark and gluons plus gravitons.
The experimental signal in this case will be jets plus missing energy. Second,
one can have the case when the decays due to mass splitting between the
first level KK excitations take place first. Then, the KK excitations
of quarks and gluons will decay to the LKP (the $\gamma^*$), radiating
low $p_T$ quarks and leptons in the process; the LKP will then decay
gravitationally, leaving behind high $p_T$ photons and gravitons (which
will appear as missing energy). Finally, one can have the intermediate
case, when the gravitational and strong/electroweak decay widths are of
comparable magnitude. Then it is possible for a $q^*$ to follow just
several steps in the decay chains ($\ref{chain}$), for example to a $l^*$,
and the KK exictation of the lepton to decay gravitationally, leaving 
behind a high $p_T$ lepton.

Which one of this scenarios will happen in practice depends on the parameters
of the model. One can easily imagine situations in which either case happens.
For example, if $N = 2$, $M_D$ is small, 
and/or the mass of the KK excitations is somewhat large,
the gravitational decay widths will tend to dominate. On the other hand,
if $N = 6$, $M_D$ is large, 
and/or the masses of KK excitations are relatively light, the
strsong/electroweak decays to the LKP will take place first. For illustration,
we present in Fig. \ref{split} the  contour lines in the $(M_D, M)$ plane for
which the KK fermion gravitational decay width is equal to the 
$l^* \rightarrow l \gamma^*$ decay width (evaluated for
$\Lambda R = 20$ ).
This means that for values
of $M_D, M$ which fall bellow the lines in the plot,
the decay to $\gamma^*$ happens
first. For points which are right on the lines (or close to them), typically
partial decays to $Z^*$ or $l^*$ happen, followed by the gravitational
decays of these excitations. For points significantly above the lines (of
order 100 GeV in $M$), the quark or gluon excitations decay directly
to gravitons and the SM partners.

\begin{figure}[t!] 
\centerline{
   \includegraphics[height=3.in]{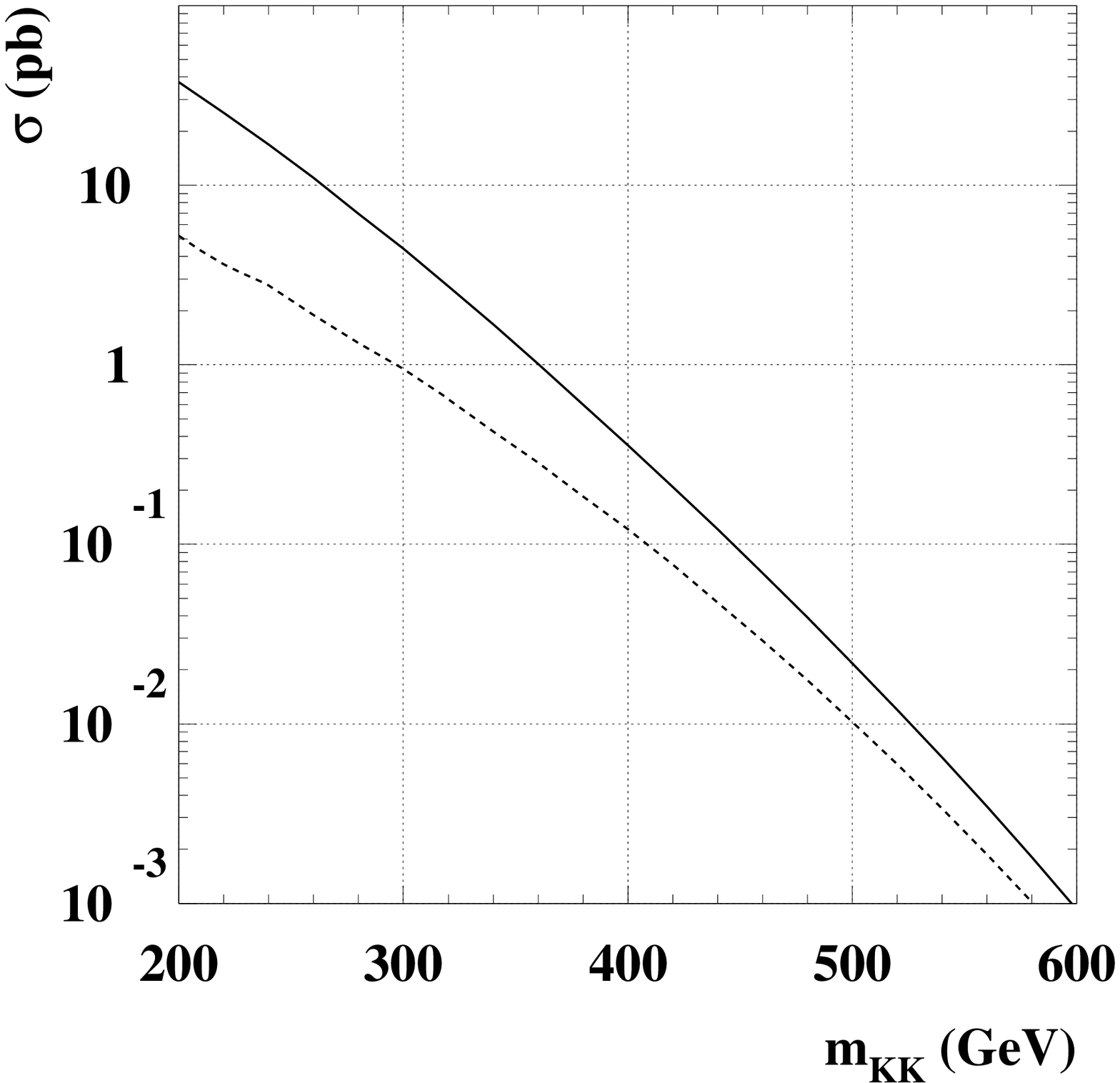}
   \includegraphics[height=3.in]{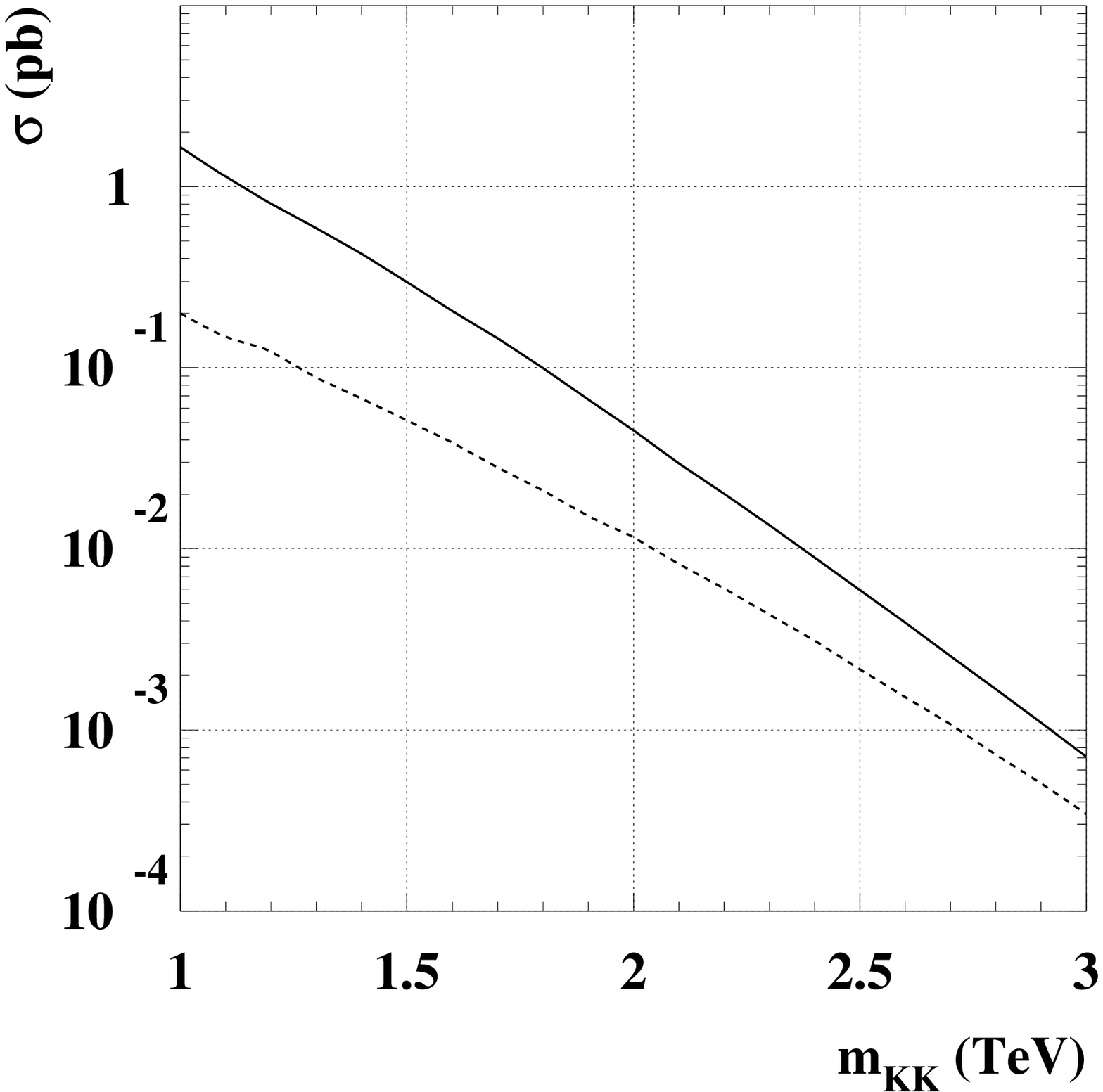}
   }
\caption{ The cross-section for dijet + missing energy production as
a function of KK excitations mass for
Tevatron Run II (left panel) and LHC (right panel). The solid lines
correspond to $N=2$ extra dimensions, and the dashed lines to $N=6$. 
The kinematic cuts applied
are described in the text.}
\label{kk_prod_pT}
\end{figure}

We will start by discussing the phenomenology of first type scenarios (with
gravitational decay widths dominant). This has been studied in some
detail in Ref. (\refcite{cmn}); here we will review the results. The mechanism
through which KK excitations are produced is the pair production processes
discussed in section {\bf 2.2}. 
The observable signal will be
two jets plus missing energy. The cross sections for this signal (with 
a $p_T$ cut on the jet transverse momentum) are shown in Fig.
\ref{kk_prod_pT}
as a function of the mass of quark KK excitations $m_{KK}$ (as shown
in Figs 1, 2, the final state contains mostly $q^*$).
The left panel corresponds to the Tevatron Run II case (with
$p_T > 100$ GeV), while the right panel corresponds to
the LHC case (with $p_T > 600$ GeV). Additional cuts are applied on
the rapidity of the individual jets $|y| < 2.5$, 
and the angular separation between the two jets $R = \sqrt{\Delta \phi^2
+ \Delta \eta ^2} > 0.4$.

The reason for applying such high $p_T$ cuts is to help eliminate the Standard
Model backgrounds. Such backgrounds will be due to the 
production of the $Z$ gauge
boson  with two jets (where $Z$ decays to $\nu \bar{\nu}$ or $\tau \bar{\tau}$
pairs),   W + 2 jets (with the
lepton from the W decay unidentifiable), $t \bar{t}$ production, with
one top decaying semileptonically to $b \bar{\nu} l$ 
with the lepton unidentified, 
and to QCD multijet production with mismeasured
missing energy ($\not{E_T}$). It is also desirable to impose an
$\not{E_T}$ cut on our signal. Since the jets we observe in our model
result from the decay of  heavy particles (the KK excitations),
they are likely to have a large $p_T$. Also, since the gravitons
have large momentum, the missing energy is likely to be large, too.
At large $ p_T, \not{E_T}$ cuts, the dominant SM background process
will come from the $Z$ + 2 jets production, and this falls rapidly 
with increasing $p_T, \not{E_T}$ (see, for example, Refs. 
(\refcite{Bityukov,tata_lhc,Gaines})). It was shown in Ref. (\refcite{cmn})
that it typically possible to separate the signal from the background
by using cuts of the type $p_T > p_T^0, \not{E_T} > 2 p_T^0$, where values
for $p_T^0$ can be chosen such as to maximize the significance, defined as
the signal divided by the square root of the background.

Another interesting issue is how could one measure the mass of KK excitations
and the number of extra dimensions $N$ if such signals are observed. Of course,
the magnitude of the cross section will give a first order approximation
for the mass of the KK excitations. However, as it can be seen in Fig. 
\ref{kk_prod_pT}, the observable cross section
also has a somewhat weaker dependence on $N$.
 Additional information about the $M,N$ parameters can be obtained by
looking at the dependence of the signal cross-section on the $p_T$ cut, and
also on the missing energy. As shown in Ref. (\refcite{cmn}), the cross
section decreases faster as a function of  $p_T$ cut for more extra dimensions;
also the missing energy is typically smaller. The reason for this behavior
is that the larger the number of extra dimensions, the higher is the mass
of the gravitons which are radiated in the decay of the KK excitations (as
discussed in the previous section); therefore, the smaller  the energy
available for the SM quarks or gluons. Analysis of such distributions
could then provide sufficient information to infer the value of the mass
$M$ precisely, as well as the number of extra dimensions. 

Finally, one should consider ways to differentiate between different
theories which give rise to  similar signals. One other obvious candidate is
supersymmetry, in which case jets + missing energy signal would arise
from gluinos (or squarks) which decay to a quark antiquark pair (or
a single quark) and a neutralino LSP (lightest supersymmetric particle). 
 Then one would see jets in the detector generated by 
these quarks, which have generally a large
energy/transverse momentum (since  the mass splitting betwen the squarks and
the LSP is typically large), while the LSP\footnote{The gluino/squarks
can also decay first to one of the other neutralinos, which in
turn may escape the detector before interaction, or decay to the LSP. 
However, even in this last case, the leptons/photons radiated during
the decay might be soft enough that they will be lost in the background.},
which is stable, will show as missing energy 
(playing the role of the graviton). An analysis aiming to 
 discriminate between
the signatures of UED/supersymmetry by using the kinematic features 
of the observable jets is underway\cite{Marius3}. 

We turn now to the analysis of the case when the decay modes allowed
by mass splitting among the first level KK excitations are dominant (this
can happen for large value of the fundamental scale $M_D$, for example).
In this scenario, as discussed in Ref. (\refcite{cmn2}), the KK excitations
of quarks and gluons pair-produced at a hadron collider 
will first decay to the LKP (the $\gamma^*$), radiating
low $p_T$ quaks and leptons in the process; the LKP will then decay
gravitationally. The signal for such a case will then be two high $p_T$
photons, accompanied by several jets and leptons with low $p_T$, and large
missing energy. 

\begin{figure}[t!] 
\centerline{
   \includegraphics[height=3.in]{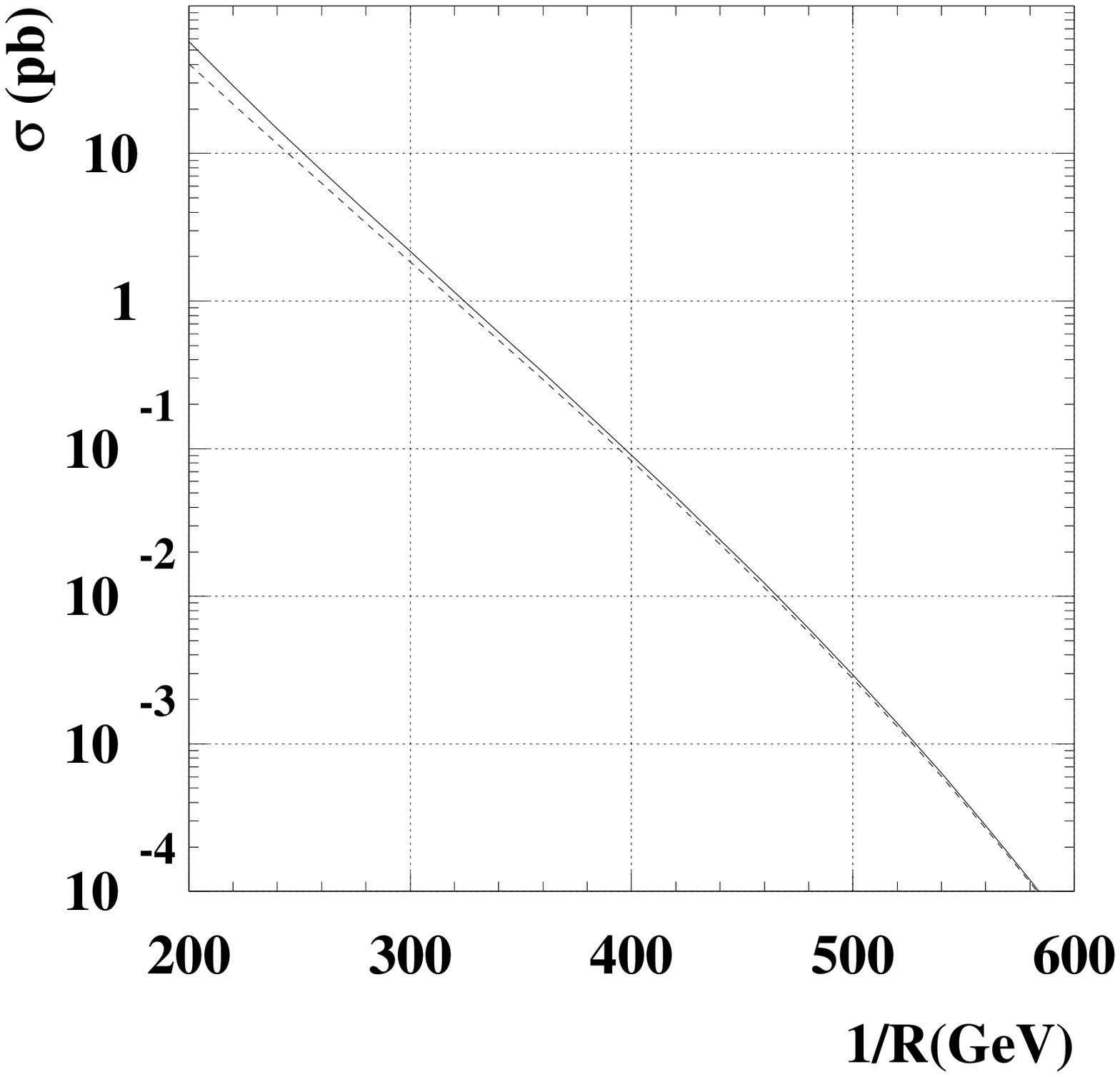}
   \includegraphics[height=3.in]{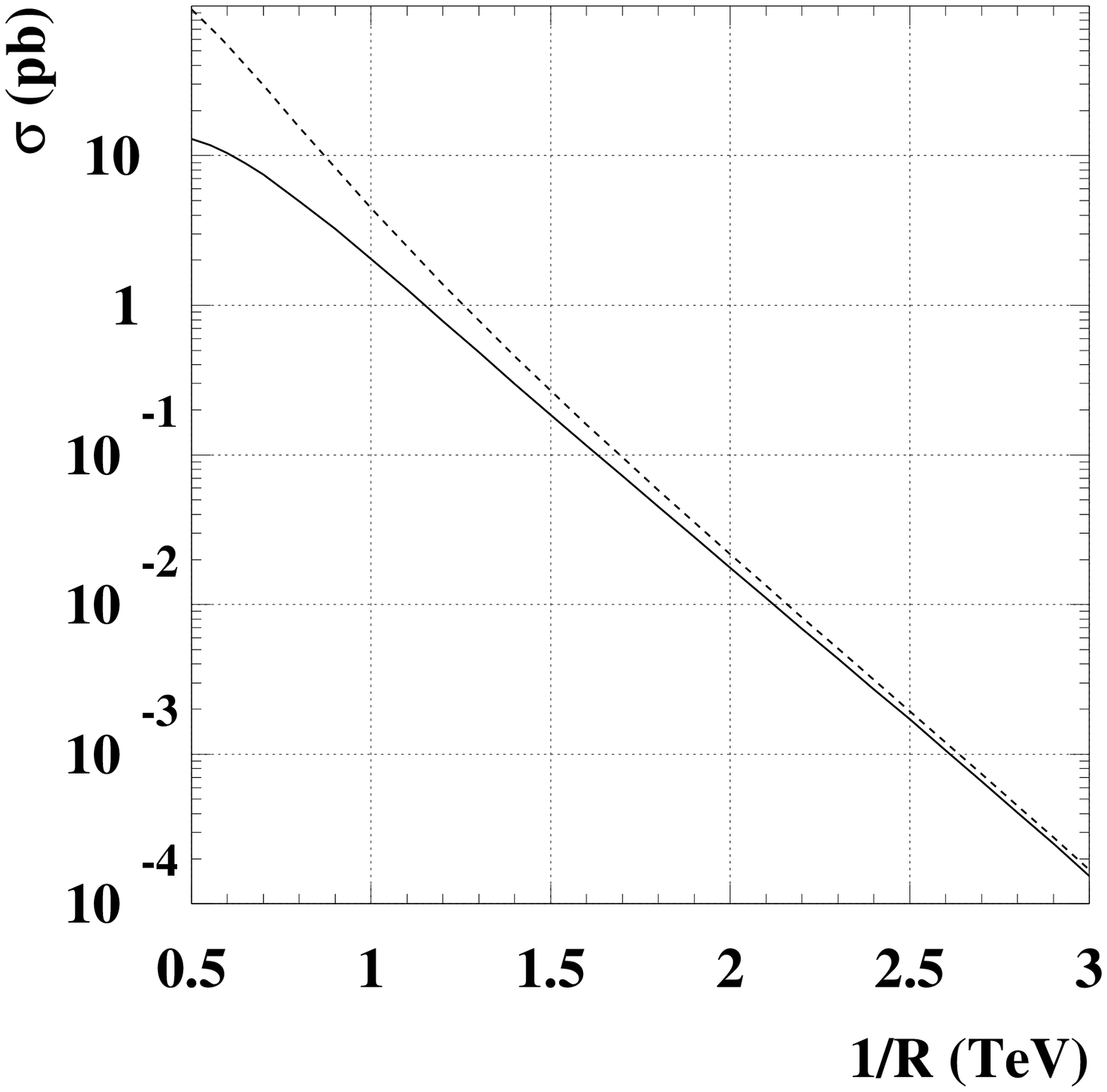}
   }
\caption{ The cross-section for diphoton + missing energy production as
a function of $1/R$ for
Tevatron Run II (left panel) and LHC (right panel). The solid lines
correspond to $N=2$ extra dimensions, and the dashed lines to $N=6$. 
The kinematic cuts applied
are described in the text.}
\label{kk_phot_pT}
\end{figure}

The Standard Model background for this signal is very small; the most
important component arise from misidentification of jets or leptons as photons,
and/or mismeasured  $\not{E_T}$. Hence, one does not need  to impose
such high $p_T$ cuts on the momenta of the observable photons. In Fig.
\ref{kk_phot_pT} we show the cross section for this signal at the
Tevatron Run II (left panel) and LHC (right panel), with the following
cuts: $p_T > 20$ GeV,  $\not{E_T} > 50 $ GeV for Tevatron, 
and $p_T > 200$ GeV,  $\not{E_T} > 200 $ GeV for the LHC. The estimated
backgrounds with these cuts are 0.5 fb at the Tevatron\cite{2grunII},
and 0.05 fb at the LHC\cite{2gtata}.
Note that these plots are shown as a function of the thickness of the brane
$1/R$, and that the masses of the KK excitations are somewhat different
from this due to radiative corrections.

Such signals can also arise in different theories, for example in a
supersymmetric model with gauge mediated SUSY breaking. In such a case, the
LSP is the goldstino/gravitino, which is
esentially massless\cite{GMSB-exp}.
 The next-to-lightest supersymmetric particle (NLSP)
may very well be a (bino-like) neutralino, which will decay to the goldstino
with the radiation of a photon. Then the squarks and gluinos predominatly
produced at a hadron collider will decay first to the NLSP (while
radiating jets and leptons which may be hard or soft, depending on the
mass splittings and the parameters of the model), and
this in turn  will decay to a hard $p_T$ photon and an invisible goldstino
(missing energy). The signal in this case may be very similar to the one 
discussed for the UED model in which the KK excitations decay first to the
LKP, and
an analysis to try to differentiate these two scenario needs to be done. 

Finally, we shall make some comments on the case when the decay widths due
to gravitational interactions and the decay widths due to mass splitting are
of the same order of magnitude. Then, one of the KK excitations of quarks
and gluons can decay gravitationally, while the other may decay first to the
LKP. The signal in this case would be jet+ photon + missing energy.  
It is also possible that one (or both) of the initial KK excitations
will decay to a KK excitation of a lepton, which in turn will decay 
gravitationally, leading to signals with jet+lepton, photon + lepton and
two leptons in the final state.

\begin{figure}[t!] 
\centerline{
   \includegraphics[height=3.in]{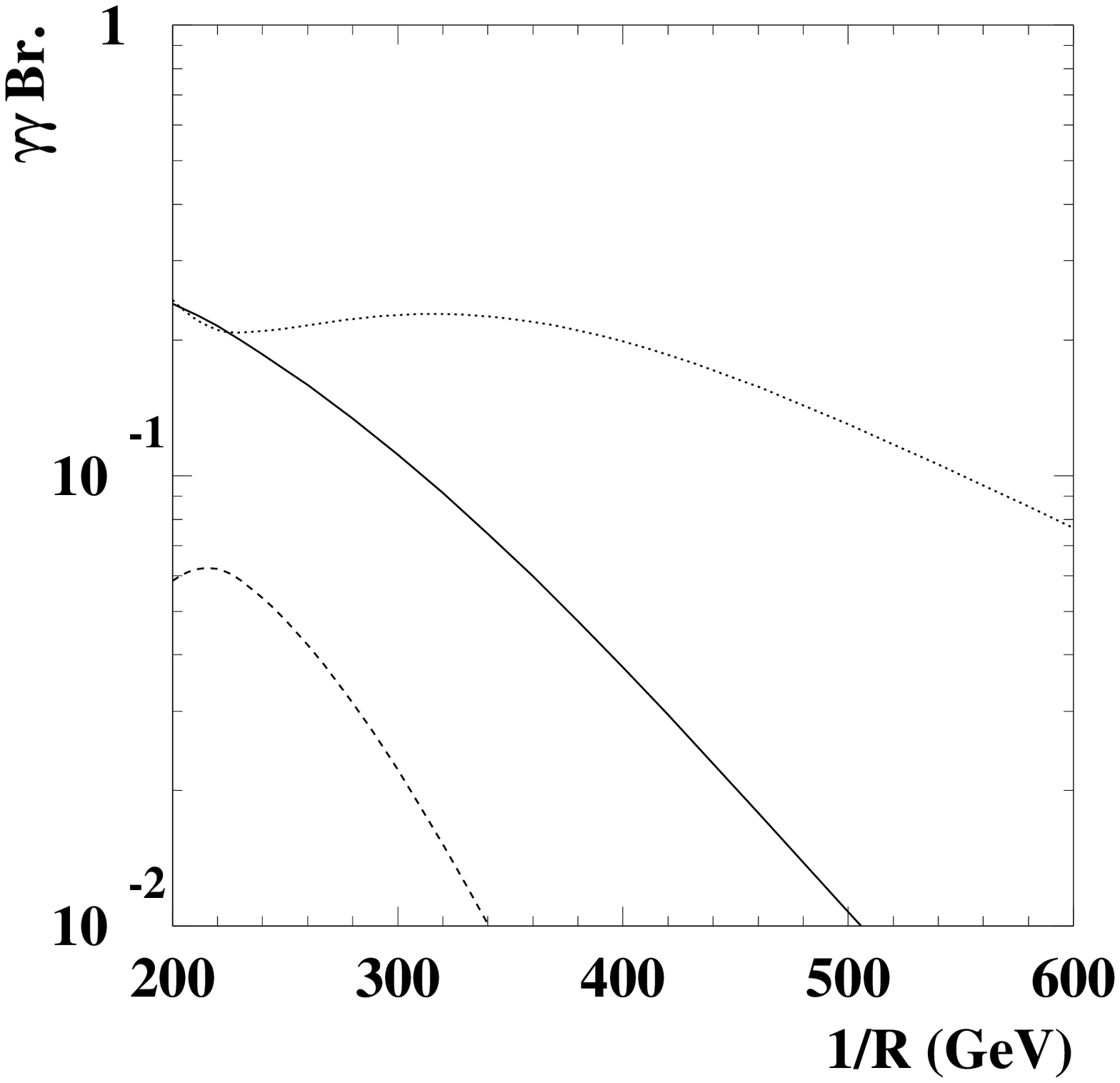}
   \includegraphics[height=3.in]{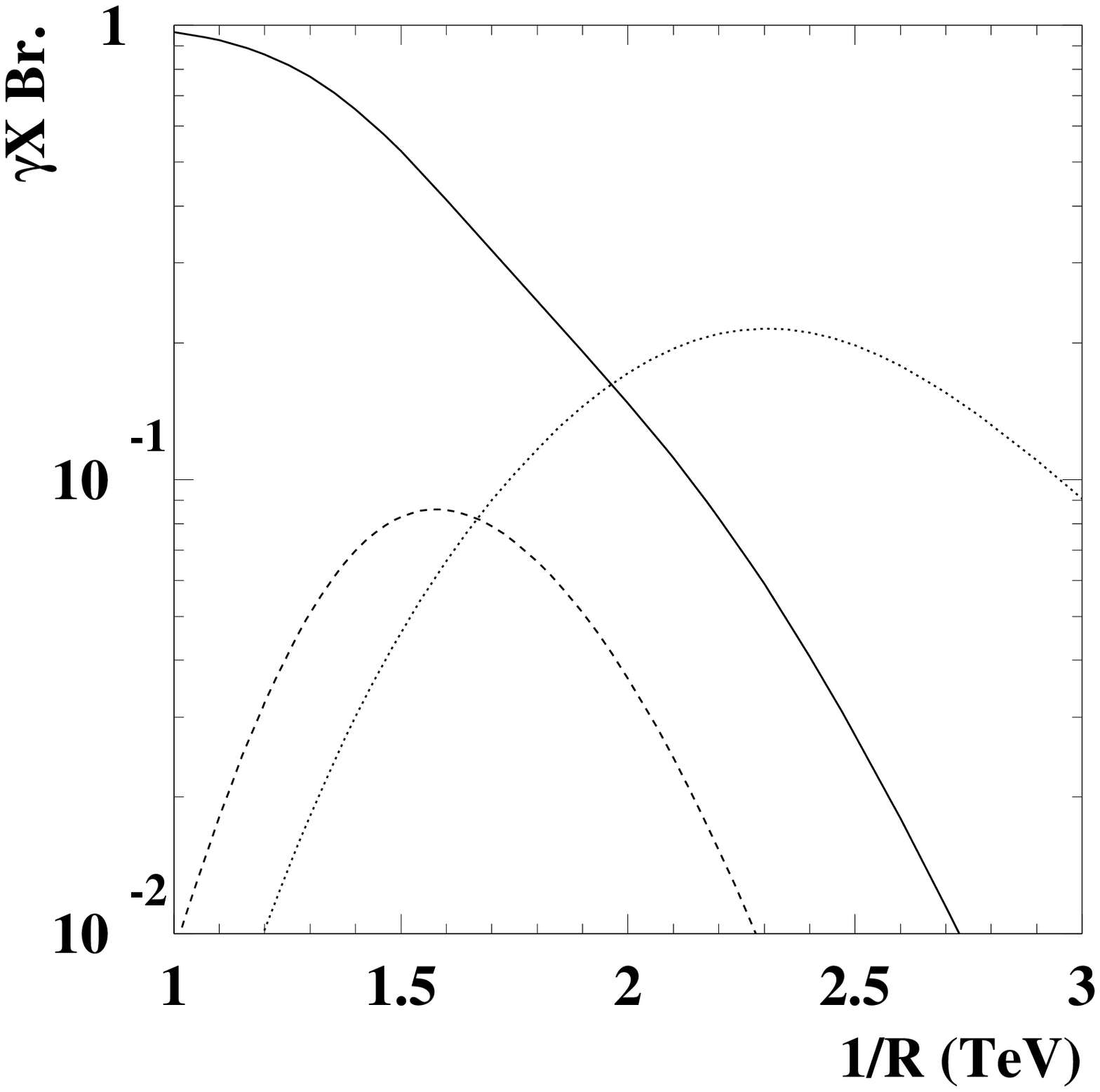}
   }
\caption{Branching ratios to final states: $\gamma \gamma$ (solid line),
jet + $\gamma$ (dotted line) and lepton ($e$ or $\mu$) + 
$\gamma$ (dashed line) for $M_D = 5$ TeV and 
$\Lambda R = 20$,
 at the Tevatron Run II and $N=2$(left), and
LHC and $N=6$ (right). (Redo the figure)}
\label{decay_modes}
\end{figure}

The relevant fact to keep in mind when discussing this case is that what will
happen is strongly dependent on the parameters of the model. Unlike
the two previously discussed cases, where the type of signal 
(as well as its magnitude) is  more or
less independend of
parameters like $M_D$ or $\Lambda$ (as long as we are in a situation 
when one or the other of the decay modes dominates), when the decay widths 
are of the same order of magnitude, the type of
signal is strongly dependent on  $M_D, \Lambda$ as well as $1/R$. For
an example of such a situation, one can look at the case when $M_D = 5$ TeV,
$\Lambda R = 20$; as can be seen in Fig. 3 in Ref. (\refcite{cmn2}),
the decay widths are of the same order of magnitude for N=2 and small
values of $1/R$ (of order 500 GeV), or, conversely,
for $N=6$ and larger values of $1/R$ (of order 3 TeV). Then, as illustrated
in Fig. \ref{decay_modes}\footnote{Note that the branching rations to
final state photons
shown here are somewhat smaller than those presented in the 
corresponding figure in Ref. (\refcite{cmn2}). This is due to the
fact that the gravitational decay widths are in fact somewhat bigger than the
estimated values used in Refs. (\refcite{cmn,cmn2}).},
 the type of signal one sees depends on 
the value of $1/R$. Moreover, there are regions in the parameter space 
where one will see different signals, if the branching ratios
for decays leading to different final states 
are of the same order of magnitude.

\subsection{Single KK excitation production}

We turn now to a discussion of the consequences which the introduction 
of a KK number violating gravitational interaction
has  on the production of matter KK excitations at colliders.
Such interaction gives rise to processes with only one KK particle in the
final state. KK gravitons may appear either as intermediate (virtual) particles
mediating the production of a SM quark/gluon and one KK excitation,
or as real particles in the final state. Since the minimum center-of-mass 
energy required
for such processes is lower than for the case of KK pair production, one can
probe higher values for $m_{KK}$ in this channel. However, since the 
gravitational interaction is involved in production, one also typically
needs a low value for the fundamental gravity scale parameter $M_D$.

\subsubsection{Gravity-mediated production}

\begin{figure}[t!] 
\centerline{
   \includegraphics[]{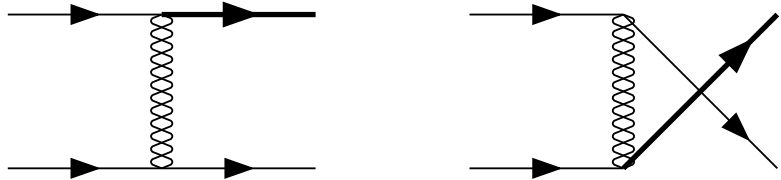}
   }
\caption{ Feynman diagrams ($t$ and $u$ channel) contributing to the
production of a $q q^*$ final state.
 The thick lines denotes the $q^*$ excitation,
while the double helix ones denotes the exchanged graviton. }
\label{fey_diag}
\end{figure}

We discuss in this section the production of a single KK excitation of
matter (quark or gluon) mediated by virtual gravitons.
The Feynman diagrams of the processes
contributing to this signal are of the type shown in Fig.  \ref{fey_diag}
(processes with gluons or $q \bar{q}$ quark pairs in the
initial and final state also have an $s$-channel
contribution). The list of all the processes with
final state $q^*$'s or $g^*$'s can be found in Ref. 
(\refcite{marius1}), together with the corresponding amplitudes.

Let us comment briefly on the graviton propagator. A single graviton
couples to matter with a strength of order $E/M_{Pl}$ (where $E$ is
the energy scale of the process under consideration). Therefore the
contribution given by a single graviton to a process as in Fig. \ref{fey_diag} 
is negligible. However, as is usual in an extra-dimensional scenario,
there is an entire tower of gravitons $h_{\vn}$ which may
contribute, and when one sums the amplitudes coming from the individual
excitations, one obtains a sizable contribution.

In the evaluation of the amplitudes for the processes of interest to us, one
therefore uses a resummed graviton propagator
\beq{grv_sum}
 D_{\mu \nu,\rho \sig}(k^2) \ = \ 
 \sum_\vn D_{\mu \nu,\rho \sig}^\vn (k^2) \ = \ 
 \kappa^2 \ \sum_\vn {\cal F}_{00|n_5} \ {i B_{\mu \nu,\rho \sig}(k)
   \over s - m_\vn^2 } \
({\cal F}^c_{10|n_5})^* \ , \eeq
where $B_{\mu \nu,\rho \sig}(k)$ is the denominator of  
the propagator for a massive spin-two particle (see, for example, Ref.
(\refcite{HLZ})), and 
${\cal F}_{00|n_5}$ and ${\cal F}^c_{10|n_5}$ are form factors
describing the interaction of the gravitons with the matter excitations
on the brane (see section 3.2). The `$00$' and `$10$' indices show 
that the graviton couples
to two Standard Model particles at one end, and to one SM particle and its 
first level KK excitation at the other end.
Terms $\sim k^{\mu} T^0_{\mu \nu}$, where $T^0_{\mu \nu}$ is the 
energy-momentum tensor associated with Standard Model matter are zero; hence
the resummed graviton propagator will have the form
$$
D(k)_{\mu \nu,\rho \sig} = \left(
\eta_{\mu \rho}\eta_{\nu \sig} +
\eta_{\mu \sig}\eta_{\nu \rho} -
{2 \over 3} \eta_{\mu \nu}\eta_{\rho \sig} \right) D(k^2) \ .
$$ 
The function $D(k^2)$ can be evaluated by replacing the sum in Eq.
(\ref{grv_sum}) by an integral (as described in section 3.3). 
Generally, this integral has to be performed numerically; however,
in the limit 
where $k^2, M$ are much smaller than the maximum mass of the gravitons, 
one can obtain the approximate expression\cite{Mitov} (valid for $N>3$): 
\beq{grv_app}
D(k^2) \ \simeq \ 
V_{N-1} \
{ 32 \over N-3}  {M\ M_S^{N-3} \over M_D^{N+2}}
{2 \sqrt{2} \over \pi^2} \
\int_0^{\pi M_s/M}  {\sin x \over 1-x^2/\pi^2} dx  \ ,
\eeq
where $M = 1/R$, and $V_{N-1}$ is the area of a sphere in $N-1$ dimensions
(due to the form factor, the integrand is not symmetric under rotations
in $N$ dimensions, but rather in $N-1$ dimensions)\footnote{We correct
the expression appearing in Ref. (\refcite{Mitov}) for the
resummed propagator by a factor of 2.}.

The quantity $M_S$ in the above expression stands for the upper limit
on the graviton masses. Note that the sum (\ref{grv_sum}) is not convergent
for $N>2$; one therefore has to impose a cut-off on the massive graviton
contributions. The scale of the cut-off is typically taken to be the same
as the fundamental gravity scale: $M_S \sim M_D$; the reason for this being 
that the scattering amplitudes we compute are valid in the low energy 
limit $E \ll M_D$. Once we get close to the gravity scale, our perturbative
field-theory description is quite possible not valid anymore, and one may
have to employ alternative descriptions, like string theory/black
hole scattering\cite{peskin-string,black-hole}.
Different choices for $M_S$ can then be thought of 
as parametrization of this new physics. In our following discussion, we will
take $M_S = M_D$, but one should keep in mind that for this type of process,
for $N>3$ the magnitude of the cross section 
varies with $M_S$ (like $M_S^{2(N-3)}$, in fact), and
the  computed signal can easily be larger or smaller depending 
upon this choice\footnote{This does not happen for the other processes
under consideration in this article. For the case of pair production
of KK excitations, the gravity interaction does not noticeably affect the 
production cross-section,
while for single KK production with a graviton, the contribution of
higher mass states is constrained by the available energy.}.
 
The observable signal for such a process will be two jets plus missing energy
(assuming that the gravitational decay width dominates). This is similar
to the case of KK pair production; however, the jets are asymmetric in this
case (since only one arises from the decay of a heavy KK particle, while
the other is produced directly).  Typically, the jet coming from the decay
of the  KK particle has higher $p_T$ (depending on the mass of the excitation),
and it should be possible to differentiate between the two. Also, due
to the fact that only one massive particle is in the final state, it
should be possible to probe higher values of $1/R$ than in the pair
production case. This is however, dependent upon the condition that the
fundamental gravity scale is low enough; since the cross-section behaves like
$1/M_D^{10}$, increasing $M_D$ will rapidly make the signal unobservable. 

\begin{figure}[t!] 
\centerline{
   \includegraphics[height=3.in]{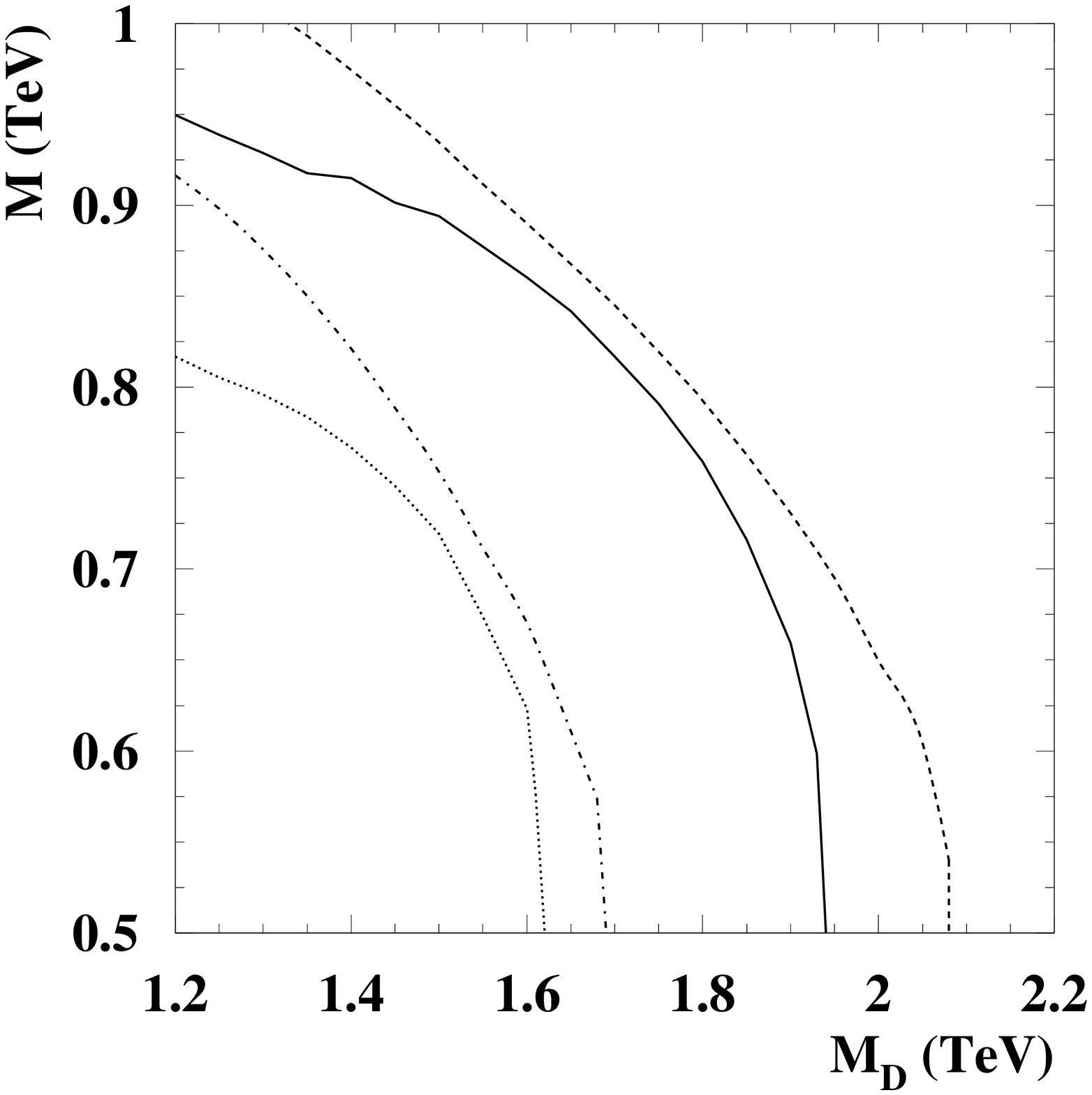}
   \includegraphics[height=3.in]{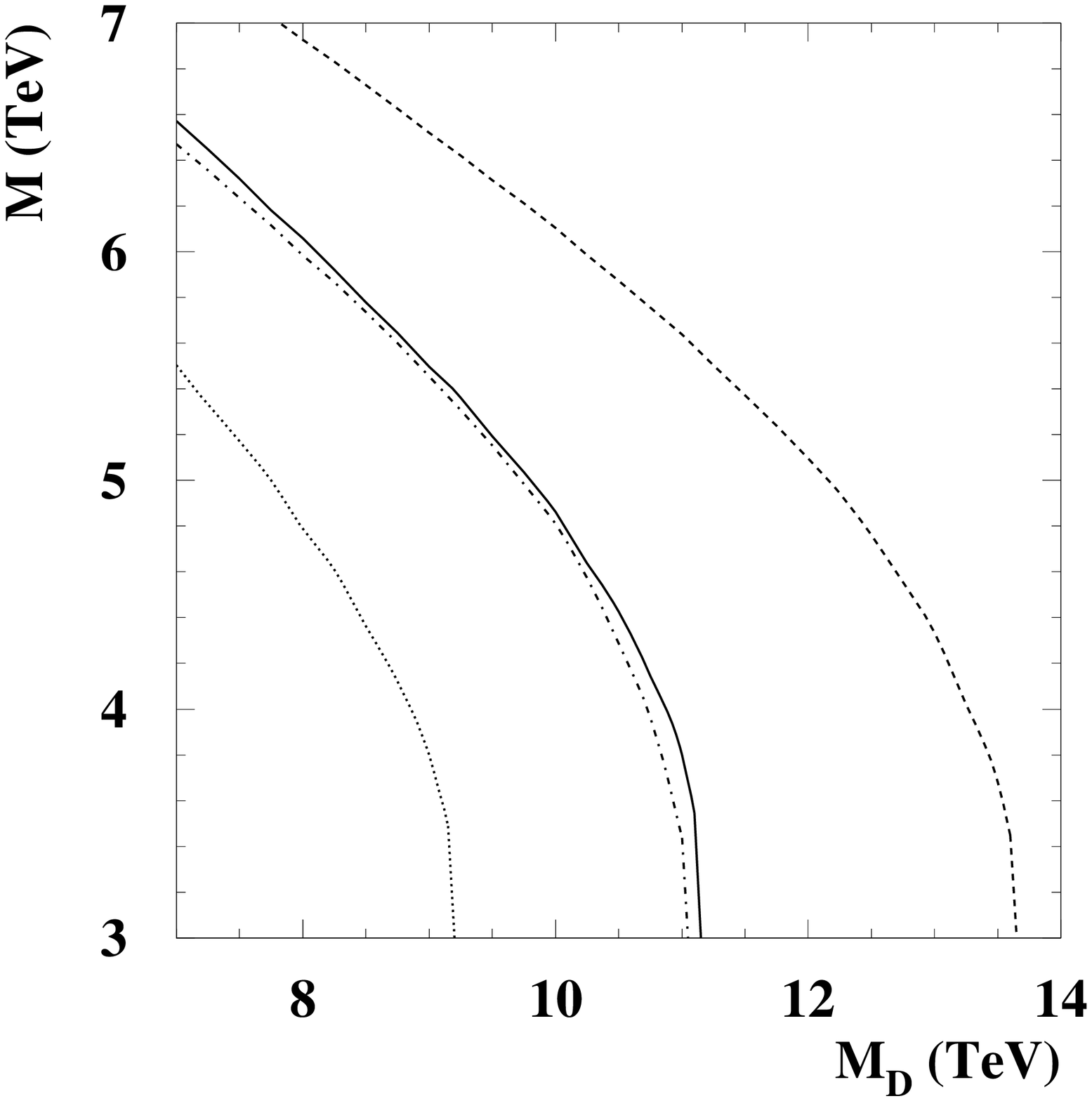}   
   }
\caption{Tevatron and LHC reach for 10 and 50 signal events (at the Tevatron), 
and  20 and 100 signal events (at the LHC). Straight and dotted lines 
corresponds to $N = 6$, and dashed and dotdashed line correspond to $N = 2$,
respectively. The cuts are described in the text.}
\label{lhc_reach}
\end{figure}

For illustration, we show in Fig. \ref{lhc_reach} 
the Tevatron and LHC  reach as a
function of $(M_D, M=1/R)$. At Tevatron, the straight and
dashed lines correspond to
a cross-section of 10 fb (with $N = 2$ and $N=6$, respectively), and
dotted and dash-dotted lines correspond to a  cross-section of 50 fb
(for $N = 2$ and $N=6$). The cuts used are $p_T > 150 $Gev for both jets,
and $\not{E_T} >$ 300 GeV. The background is the same as the one discussed
for the pair production case, and with an integrated luminosity of 
2 fb$^{-1}$, it amounts to one event (the signal being then 10 and 50
events). We see that for relatively low values of $M_D$, one is able 
to almost double the $m_{KK}$ discovery reach, from four to five hundred GeV
at the Tevatron (Fig \ref{kk_prod_pT}), to almost 800 GeV.

The right panel in Fig. \ref{lhc_reach} shows the LHC reach. 
The straight and
dashed lines correspond to
a cross-section of 0.2 fb (with $N = 2$ and $N=6$, respectively), and
dotted and dash-dotted lines correspond to a  cross-section of 1 fb
(for $N = 2$ and $N=6$). The cuts used are $p_T > 800 $ Gev for both jets,
and $\not{E_T} >$ 1.6 TeV. The background with these cuts (and 100 fb$^{-1}$
integrated luminosity) is around 10 events, while the signal will be
20, respectively 100 events. The discovery reach  for low values of
$M_D$ also increases in this
case compared to the pair production case.

This type of signals have been studied in Ref. (\refcite{marius1}). However,
one can also have $l l^*$ and $\gamma \gamma^*$ final states, from 
processes with the exchange of $s$ channel gravitons. This requires
that the initial state is either $q \bar{q}$ or $g g$, therefore
the production cross section will be somewhat smaller than for the case
of quarks of gluons in the final state.  However, the
observable signal will be two high $p_T$ photons or leptons,
and the SM background will be much reduced. Detailed simulations have 
not been performed yet, but one would expect that the 
discovery reach in this channel will  be as large 
as in the two jets case, or even larger.

Finally, it is interesting to consider
that  in searching for universal extra dimensions,
one can look at KK pair production and gravity mediated
single KK production as somewhat complementary channels. If $M_D$ is
relatively small, the pair produced KK excitations will decay to 
jets + gravitons, thus making them somewhat hard to see at hadron colliders;
however, in this case the cross section for single KK production can be quite
large. On the other hand, if the gravity scale is larger, the cross-section
for the production of a single KK excitation will be small; but then 
the pair-produced KK excitations will decay first to $\gamma^*$, and
the two photons + large  $\not{E_T} $ signal will make the signal in this
channel easier to see at hadron colliders.

\subsubsection{Final state gravitons}

We end our review of the phenomenology of universal extra dimensions
with a short discussion of the case when a single KK excitation of matter is
produced with a graviton KK excitation in the final state. Again, the production
rate for a particular graviton excitation is small (of order $E/M_{Pl}$), but
when summing over the KK tower one can obtain a sizable cross-section.

The amplitudes contributing to the processes of interest here are the same 
ones encountered in the evaluation of the production rate for a Standard
Model particle plus a KK graviton in the usual ADD model\cite{peskin-add,GRW}.
Indeed, where one produces the KK excitation of a quark or gluon, one can
also produce the Standard Model particle together with the graviton. If the
KK quark or gluon decays to a jet, one obtains the same type of signal (jet
plus missing energy) in both cases. Since the process with the SM particle
in the final state is phase-space favored (no need to produce a massive 
excitation), one can conclude that the production of a KK excitation of matter
is at most a small correction.


 However, a more detailed analysis is needed to support this conclusion.
As discussed above, high $p_T$ and missing energy cuts (of order of TeV
at the LHC) might be needed to separate the signal in this case from the
SM background. Since the jet coming from the decay of a heavy KK excitation
will typically have a higher transverse momentum than a jet coming
from a directly produced quark or gluon, it is possible that in the
region of phase space of interest experimentally the signals for the
two types of processes are of comparable orders of magnitude. 
Moreover,
 if one sees just an excess of jets with missing energy at a hadron
collider, while it is possible to infer the existence of new physics,
it will not be clear what type of physics it is. For example, besides
the fat brane scenario we discuss here, 
 it could be supersymmetry, 
or a pure ADD model (with matter restricted on the 4D brane). An analysis
of relevant phenomenological observables (like $p_T$  
distributions)  will be needed to determine
the exact theory.

The signal for this case has been studied in
Ref. (\refcite{Marius2}). Results
indicate that the cross-section for the production of gravitons + KK 
excitations of matter is indeed significantly smaller than the cross section
for the production of gravitons + SM matter. The phase space constraints
play a role, but additionally this is due in part to the
effect of the form-factor. As discussed in section 3.2, the interaction
vertex of matter with gravity aquires a form factor ${\cal F}^c_{1|n_5}(x)$,
$x =  2 \pi n_5 R/ (l r)$,
proportional to the superposition of wave functions of matter and gravity 
in the fifth direction (see Eq. (\ref{dec_ff1})).
Since $|{\cal F}^c_{1|n_5}(x)|^2 \sim x^2$ for small $x$, this has as result
a suppression of the production cross section\footnote{One might think
that this factor should not be important for values of $N$ greater than 2,
given that most of the production cross-section will then be due to radiation
of heavy gravitons. However, even if most of the cross section correspons to
large $n = \sqrt{\vn^2}$, it also corresponds to small $n_5$. This can be 
understood by noting that most of the volume of a $N$ dimensional
sphere of unit radius comes from relatively small values for any one of 
the coordinates.}.

As a consequence, it will be quite hard to see the production of a matter
KK excitation with a graviton in the jet plus missing energy channel.
However, a better chance for the observation of this process 
happens for the case
when the mass-splitting decay modes dominate, and the gluon or quark 
excitation decays first to $\gamma^*$. The background for the signal
in this case
(high energy photon with missing transverse energy) is low in the Standard
Model, as well as in the case of production of KK gravitons with a SM photon
(since the strength of the interaction is governed by the electroweak coupling),
and it is possible to measure a small signal cross-section.
Moreover, for small values of $N$, the production cross section decreases
slower with $M_D$ than for the case of gravity-mediated single KK production
($\sig \sim 1/M_D^{N+2}$ versus $1/M_D^{10}$ in the later case), so one
can potentially probe larger values of $M_D$. Thus, if the mass of the matter
KK excitations is not too big (of order  1 TeV or smaller), for $N=2$ it might
be possible to probe values for $M_D$ as large as 40 TeV.

\section{Conclusions}

Theoretical considerations (like string theory)
suggest that there may be more than 4 space-time dimensions. If the 
compactification scale for the extra dimensions is relatively large,
they may be the of relevance for low energy physics. In particular,
models where matter propagates in extra dimensions of inverse TeV scales 
have interesting implications for the phenomenology of present-day and
future colliders. We review in this article the collider
phenomenology associated
with a particular class of models, Universal Extra Dimensions, in which all
matter fields propagate in the bulk.

A characteristic feature of such theories is Kaluza Klein parity conservation,
which requires that the KK excitations of Standard Model particles are 
produced in pairs. Moreover, the lightest KK particle (LKP) is stable, and
generally weakly interacting.
The masses of the first level excitations aquire radiative corrections,
with strongly interacting particles getting higher masses than the
excitations of leptons and weakly interacting bosons. Therefore, first KK
level quarks and gluons produced at a hadron collider will decay to the LKP,
radiating semi-soft leptons and jets. Experimental signals for this scenario
will be similar to that of a supersymmetric model with almost degenerate mass
spectrum for the superpartners, and the lightest supersymmetric particle (LSP) 
being a neutralino.  

Adding gravitational interactions will modify somewhat this picture. Such 
interactions might break KK number conservation, since gravity and matter do not
have to propagate on the same scale in the extra dimensions. In fact, there are
theoretical arguments supporting the view that gravity should propagate in 
'large' extra dimensions with inverse eV size, while matter fields
 should be stuck 
close to the 4D brane, with only a length $\sim 1/M_D \sim $ TeV$^{-1}$ 
accesible to them (the fat brane scenario). Then the KK excitations of matter
(including the LKP) can decay to SM particles and KK gravitons.

The phenomenology resulting from KK pair production will be then quite 
different. Generally speaking, there will be two high $p_T$ particles/jets 
in the
final state, corresponding to the decay of the KK excitations, and missing
energy associated with gravitons. The exact nature of the final state particles
depends on the details of the model; thus, if gravitational decays widths are
large, the KK excitations of quarks and gluons produced at a
hadron collider  will decay directly to SM quarks and gluons, which will appear
as jets in the detector. However, if the gravitational decays widths are small,
the KK quarks and gluons will first decay to the LKP (radiating soft jets and 
leptons in the process) which then will decay gravitationally, 
leading to a signal with two high $p_T$ photons in the final state. If the
two decay modes are of comparable importance, one can obtain a mix-up of
final state high $p_T$ particles, including photons, jets and leptons.

Moreover, in this scenario it is possible to produce a single KK 
excitation of matter
through processes involving gravitons. Since the effective
strength of  the gravitational interaction is $\sim 1/M_D^2$, this requires
that the fundamental gravity scale $M_D$ be rather low. However, since one
need to produce only one massive particle in the final state, one can probe
higher values for the masses of matter KK excitations in this type of process. 

The experimental signals for such production processes can also be  
two jets, photons or leptons, 
jet and photon or lepton, single jet or single photon with missing energy 
in the final state. Such signals can arise in a variety of supersymmetric 
models too; for example, final states with jets and/or leptons can arise in 
MSUGRA models with squarks and gluinos decaying to a significantly
lighter LSP neutralino (cascade decays through charginos might account for the 
leptons).  Final states with high $p_T$ photons can arise in gauge mediated
supersymmetry breaking (GWSB) models,
where the LSP is a very light goldstino/gravitino. In 
conclusion, this class of models leads
to a rich phenomenlogy, 
and a careful study is needed to discriminate between SUSY and UED signals
at  hadron colliders.

\section*{Acknowledgments}

The author wishes to thank B. Dobrescu, S. Nandi and M. Rujoiu 
for many helpful
discussions. This work is supported  in part by the US Department
of Energy grant number DE-FG02-85ER40231.

\end{document}